\documentclass[conference]{IEEEtran}
\usepackage{stmaryrd}
\usepackage{subfigure}
\usepackage{colortbl,multirow,hhline}
\usepackage[pdftex]{graphicx}
\usepackage{pifont}
\usepackage{listings}
\usepackage{amssymb,amsmath}
\usepackage{color}
\usepackage{array}
\usepackage{acronym}
\usepackage{algorithm}
\usepackage{url}
\usepackage{paralist}
\usepackage{caption}
\usepackage{algorithm}
\usepackage{algorithmicx}
\usepackage{algpseudocode}
\usepackage{tikz}
\usetikzlibrary{arrows,positioning}
	
\usepackage[acronym]{glossaries}
\usepackage{booktabs,bigstrut,rotating,multirow,hyperref,lineno}

\usepackage[usenames,dvipsnames]{pstricks}
\usepackage{epsfig}
\usepackage{pst-grad} 
\usepackage{pst-plot} 

\usepackage{graphicx}

\usepackage{todonotes}

\newacronym{rl}{RL}{reinforcement learning}

\ifCLASSINFOpdf
\else
\fi
\acrodef{flc}[FLC]{Fuzzy Logic Controller}
\acrodef{rl}[RL]{Reinforcement Learning}

\begin{document}
%

\title{Self-Learning Cloud Controllers: \\ 
\emph{Fuzzy Q-Learning for Knowledge Evolution}}




%
\author{\IEEEauthorblockN{Pooyan Jamshidi\IEEEauthorrefmark{1},
Amir Sharifloo\IEEEauthorrefmark{3},
Claus Pahl\IEEEauthorrefmark{2},
Andreas Metzger\IEEEauthorrefmark{3},
Giovani Estrada \IEEEauthorrefmark{4}
}
\IEEEauthorblockA{\IEEEauthorrefmark{1}Imperial College London, UK}
\IEEEauthorblockA{\IEEEauthorrefmark{2}IC4, Dublin City University, Ireland}
\IEEEauthorblockA{\IEEEauthorrefmark{3}University of Duisburg-Essen, Germany}
\IEEEauthorblockA{\IEEEauthorrefmark{4}Intel, Ireland\\}
}


\maketitle

\begin{abstract}
Cloud controllers aim at responding to application demands by automatically scaling the compute resources at runtime to meet performance guarantees and minimize resource costs. Existing cloud controllers often resort to scaling strategies that are codified as a set of adaptation rules.  However, for a cloud provider, applications running on top of the cloud infrastructure are more or less black-boxes, making it difficult at design time to define optimal or pre-emptive adaptation rules. Thus, the burden of taking adaptation decisions often is delegated to the cloud application. Yet, in most cases, application developers in turn have limited knowledge of the cloud infrastructure. 

In this paper, we propose learning adaptation rules during runtime. To this end, we introduce \texttt{FQL4KE}, a self-learning fuzzy cloud controller. In particular, \texttt{FQL4KE} learns and modifies fuzzy rules at runtime. The benefit is that for designing cloud controllers, we do not have to rely solely on precise design-time knowledge, which may be difficult to acquire. \texttt{FQL4KE} empowers users to specify cloud controllers by simply adjusting weights representing priorities in system goals instead of specifying complex adaptation rules. The applicability of \texttt{FQL4KE} has been experimentally assessed as part of the cloud application framework \texttt{ElasticBench}. The experimental results indicate that \texttt{FQL4KE} outperforms our previously developed fuzzy controller without learning mechanisms and the native Azure auto-scaling.


\end{abstract}


%
\IEEEpeerreviewmaketitle

\section{Introduction}
\label{sec:introduction}


The economic model of pay-per-use behind cloud computing allows companies to rent a variety of resources for a certain period and access them via Internet \cite{Jamshidi2013}. Despite the advantages, dynamic acquisition and release of resources is a big challenge for applications due to the uncertainty introduced by workload, cost and user requirements. In order to address this challenge, many different approaches \cite{Lorido-Botran2014,netto80evaluating,caron2012auto} referred as \emph{auto-scaling} has been proposed. The current state of practice relies on threshold-based rules and thanks to their simplicity and intuitiveness, they are mostly offered by many commercial cloud providers/platforms such as Amazon EC2 \cite{AmazonEC2}, Microsoft Azure \cite{azure}, OpenStack \cite{OpenStack}. The typical practice is to define a manageable, usually small and comprehensible set of scaling rules, assuming a linear and constant dependency between resource assignments and performance improvements, while in Internet scale applications, the complexity of application architecture, the interferences among components and the frequency by which hardware and software failure arise typically invalidate these assumptions \cite{Pu2013,Gambi-TSC}.

The research community has investigated many alternative approaches. There has been solutions based on classical control theory \cite{filieri2015software} and on knowledge-based controllers \cite{xu2007use} and thus suffer from similar limitations \cite{DynaQoS}. Traditional capacity planning approaches \cite{Sedaghat2013} based on queuing theory or similar model-based approaches \cite{Ardagna2014} do not fully address the dynamics of cloud applications due to the mathematical simplifications and/or their static nature since the models are complex to be evolved at runtime, often resort to parameter tuning \cite{jamshidi2013framework}. The recent trends based on self-organizing controllers have shown to be a better fit for the complexity of cloud controllers \cite{Gambi-TSC,Bodk2009}. However, a practical challenge remains unanswered, that is reliance on users for defining cloud controllers. There are some facts behind this challenge. First, from the cloud provider's perspective, the details of the applications are simply not visible, making it difficult to accurately devise optimal set of scaling rules. Thus, the burden of determining such rules falls on the cloud users, who do not have enough knowledge about the workloads, infrastructure or performance modeling.


\subsection{Research Challenges}
In our previous work \cite{jamshidi2014autonomic}, we exploited fuzzy logic to facilitate user intuitive knowledge elicitation. The key strength of fuzzy logic is the ability to translate human knowledge into a set of intuitive rules. During the design process of a fuzzy controller, a set of \texttt{IF-THEN} rules must be defined. 
Although we showed that users are more comfortable with defining auto-scaling rules using fuzzy linguistic variables \cite{jamshidi2014autonomic}, the rules have to be defined at design-time leading to the following issues:
\begin{inparaenum}[(i)]
\item Knowledge may not be available (user cannot prescribe any rule);
\item Knowledge may be available but in part (user can only specify partial rules for some situations);
\item Knowledge is not always optimal (user can specify the rules but they are not effective, e.g., redundant rules);
\item Knowledge may be precise for some rules but may be less precise (i.e., contains uncertainty) for some other rules (depending on the degrees of a priori knowledge). 
\item Knowledge may need to change at runtime (rules may be precise at design-time but may drift at runtime).
\end{inparaenum} 
As a result, user defined rules may lead to sub-optimal scaling decisions and loss of money for cloud application providers. 

\subsection{Research Contributions}
In order to address the above challenge, we develop an online learning mechanism, \texttt{FQL4KE}, to adjust and improve auto-scaling policies at runtime. More specifically, we combine fuzzy control and Fuzzy Q-Learning (FQL) \cite{Glorennec1997} in order to connect human expertise to continuous evolution machinery. Q-learning is a method that allows the system to learn from interaction with the environment, where the learning is performed via a reward mechanism \cite{Watkins1992}. The combination of fuzzy control and the Fuzzy Q-Learning proposes a powerful self-adaptive mechanism where the fuzzy control facilitates the reasoning at a higher level of abstraction (i.e., human reasoning) and the Q-learning allows to adapt/adjust the controller. 

The main \emph{contributions} of this work are as follows:
\begin{enumerate} [(i)]
  \item a self-learning fuzzy controller, \texttt{FQL4KE}, for dynamic resource allocations.
  \item a tool, \texttt{ElasticBench}, as a realization and a means for experimental evaluations of the entire approach.
\end{enumerate}
The main implication of this contribution is that we do not need to rely on the knowledge provided by the users anymore, \texttt{FQL4KE} can start adjusting application resources with no priori knowledge. This means the auto-scaling controller can start working with an empty knowledge base and obtain knowledge at runtime, through the knowledge evolution mechanism.

The rest of the paper is organized as follows. Section \ref{sec:background} gives the underlying concepts and motivates the work. Section \ref{sec:approach} describes the mechanisms that constitutes our solution followed with a realization in Section \ref{sec:realization}. Section \ref{sec:experiment} discusses the experimental results following by implications and limitations of this work in \ref{sec:discussion}. Finally, Section \ref{sec:related} discusses the related work and Section \ref{sec:future} concludes the paper.

\section{Motivation and justification}
\label{sec:background}

\subsection{Motivation}
\label{sec:policy-notion}

Dynamic resource provisioning, also called auto-scaling, is a decision making problem. 
\emph{Cloud controllers} that realize auto-scaling plays the role of a controller that observes the resource consumption of application and manipulates the provisioning plans. 
In a more formal wording, computing nodes are allocated to cloud-based applications by regularly observing the workload, $w$, in user request per time unit and the current performance, $rt$, as average response time of the application. The cloud controller decides to allocate more nodes or release some existing ones in order to keep the performance $rt$ below a desired performance $rt_{desired}$ declared by the SLA of the application while minimizing costs.

There are some common and noticeable characteristics that often challenge the existing auto-scaling techniques and tools:
\begin{inparaenum}[ (i)]
\item the environment is non-episodic, i.e., current choice will affect future actions;
\item cloud infrastructures are complex and difficult to model;
\item workloads are irregular and dynamic.
\end{inparaenum}
These characteristics of the environment in which cloud controller operates as an agent require to solve sequential decision problems, where previous actions in specific states affect future ones. The common solution types for such kind of problem is to elaborate a \emph{plan}, \emph{policy} or \emph{strategy} to act upon specific situation at hand. In this paper, we use the term \emph{policy} as the knowledge inside cloud controllers that we aim to learn at runtime. As a results, policies determine the decisions that controllers produce based on different situations (i.e., the state in which the cloud application is in).

In this setting, the choice of assigning the responsibility of allocating the required resources to the application provider depends on the nature of the applications that typically include several components implementing the application logic through complex interactions. Hence, platform provider's native auto-scaling, like the ones monitoring only system metrics, is sub-optimal with respect to application-specific solutions such as autonomic controllers. The key reason for this is that the native auto-scaling only have limited knowledge as application architecture is not fully visible for providers.

Although a variety of techniques have been proposed, developing efficient and effective auto-scaling approaches is a challenging activity\cite{Lorido-Botran2014}. While reactive techniques allocate resources according to the latest system demands \cite{Maurer2011}\cite{caron2012auto}\cite{Marshall2010}, proactive techniques forecast the future needs upon which they adjust the resource \cite{gandhi2014adaptive}\cite{lim2010automated}\cite{Shen2011}. The majority of the approaches apply threshold-based rules to trigger adaptation actions, which is highly sensitive to the selected thresholds \cite{Masoumzadeh2013}. The proactive approaches are also prone to severe prediction error as mostly use linear model to predict future situation \cite{jamshidi2014autonomic,Lorido-Botran2014}. In this work, we investigate the application of fuzzy controllers equipped with machine learning, here Q-Learning, to address the aforementioned challenges.

\subsection{Reinforcement Learning for Elasticity Decision Making}
\label{sec:rlql}

\begin{figure}
\begin{center}
\includegraphics[width=7cm]{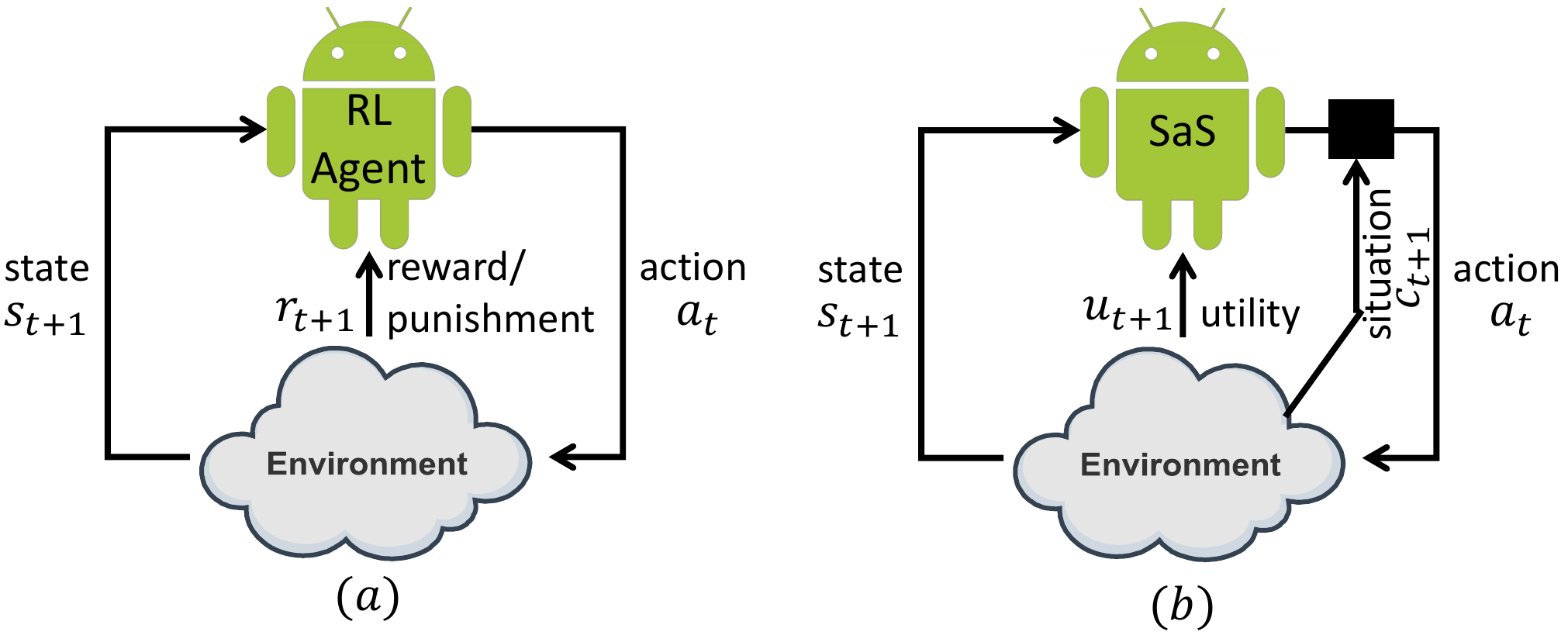}
\caption{RL agent interaction with environment.}
\label{fig:rl}
\end{center}
\end{figure}

As depicted in Figure \ref{fig:rl}(a), an agent takes action $a_i$ when the system is in state $s_t$ and leaves the system to evolve to the next state $s_{t+1}$ and observes the reinforcement signal $r_{t+1}$. The process of decision making in elastic systems can be represented as an interaction between cloud controllers and the environment. The cloud controller monitors the current state of the system through its sensors. Based on some knowledge, it chooses an action and evaluates feedback reward in the form of \textit{utility functions} \cite{Walsh2004}. 
$Situation$ allows the system to know when it must monitor the state and reward, and also when it must take the action corresponding to the state (i.e, triggers the scaling action). An elastic system may stay in the same state, but should take different actions in different situations, workload intensity. 

\begin{figure}
	\begin{center}
		\includegraphics[width=7cm]{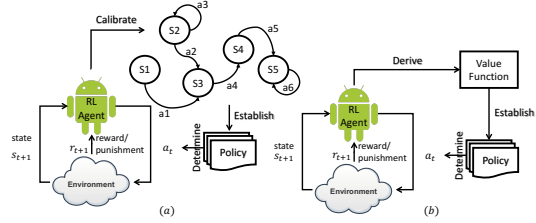}
		\caption{Model-based vs. model free RL.}
		\label{fig:model-based-free}
	\end{center}
\end{figure}

To derive an action, the agent uses a $policy$ that aims to increase the future rewards at the long run. A $model$ of the environment, can help the decision making by agent (cf. Figure \ref{fig:model-based-free}(a)); however, it is not always feasible to have such a model available in many application areas. Model-free reinforcement learning (hereafter \emph{RL}) techniques have been developed to address this issue, which are interesting for cloud computing problems due to the lack of environment models.
In this paper, we use a model-free approach. More specifically, we use Q-Learning algorithm that computes the optimal policy with regard to both immediate and delayed rewards. In this case, a cloud controller learns a value function (cf. Figure \ref{fig:model-based-free}(b)) that gives the \emph{consequent} of applying different policies. 

\subsection{Justification of RL for Dynamic Resource Provisioning}
\label{justification}

There exist a couple of techniques that might be considered as suitable candidates for the problem of dynamic resource allocation: Neural Networks (NNs), Evolutionary Algorithms, Swarm Intelligence and Reinforcement Learning. For the research we report in this paper, we had specific reasons for choosing RL amongst the alternative choices. NNs are appropriate for function approximation, classification and pattern recognition. A drawback of NNs is that they require a large diversity of pre-operation training before real operations, which can be a sever constraints as training trials in the context of cloud auto-scaling are costly and time consuming \cite{Gambi-TSC}. 

Evolutionary computation, such as genetic algorithm and particle swarm optimization, is based on randomly producing and comparing the evolution of many genes and particles, each of which represents a different configuration of a cloud controller. As a result, to evaluate the optimality of each particle, the evaluation of the corresponding controller should be carried out in many trials. The lack of generality in the definition of cloud controllers is also a constraint. Therefore, the optimization phase must be repeated.

Finally, for the following reasons, RL is an appropriate fit:
\begin{itemize}
	\item Workload for cloud-based applications are unpredictable and obtaining an actual training data set that is representative of all runtime situations becomes a mission impossible task. Unlike other supervised learning (e.g., NNs) approaches, in RL a training data set is not necessary.
	\item Due to the unpredictably of workload and complexity of the cloud-based application, providers do not have a complete knowledge to take proper scaling actions. 
\end{itemize}

\section{Fuzzy Q-Learning for Knowledge Evolution}
\label{sec:approach}

This section presents our solution \texttt{FQL4KE}. By combining fuzzy logic and Q-Learning,  \texttt{FQL4KE} deals with uncertainty caused by the incomplete \emph{knowledge}. Expert knowledge, if available, is encoded in terms of rules. The fuzzy rules are continually tuned through learning from the data collected at runtime. In case there is no (or limited) knowledge available at design-time, \texttt{FQL4KE} is still able to operate.


\subsection{FQL4KE Building Blocks}
\label{fqlk-architecture}

\begin{figure}
\begin{center}
\includegraphics[width=9cm]{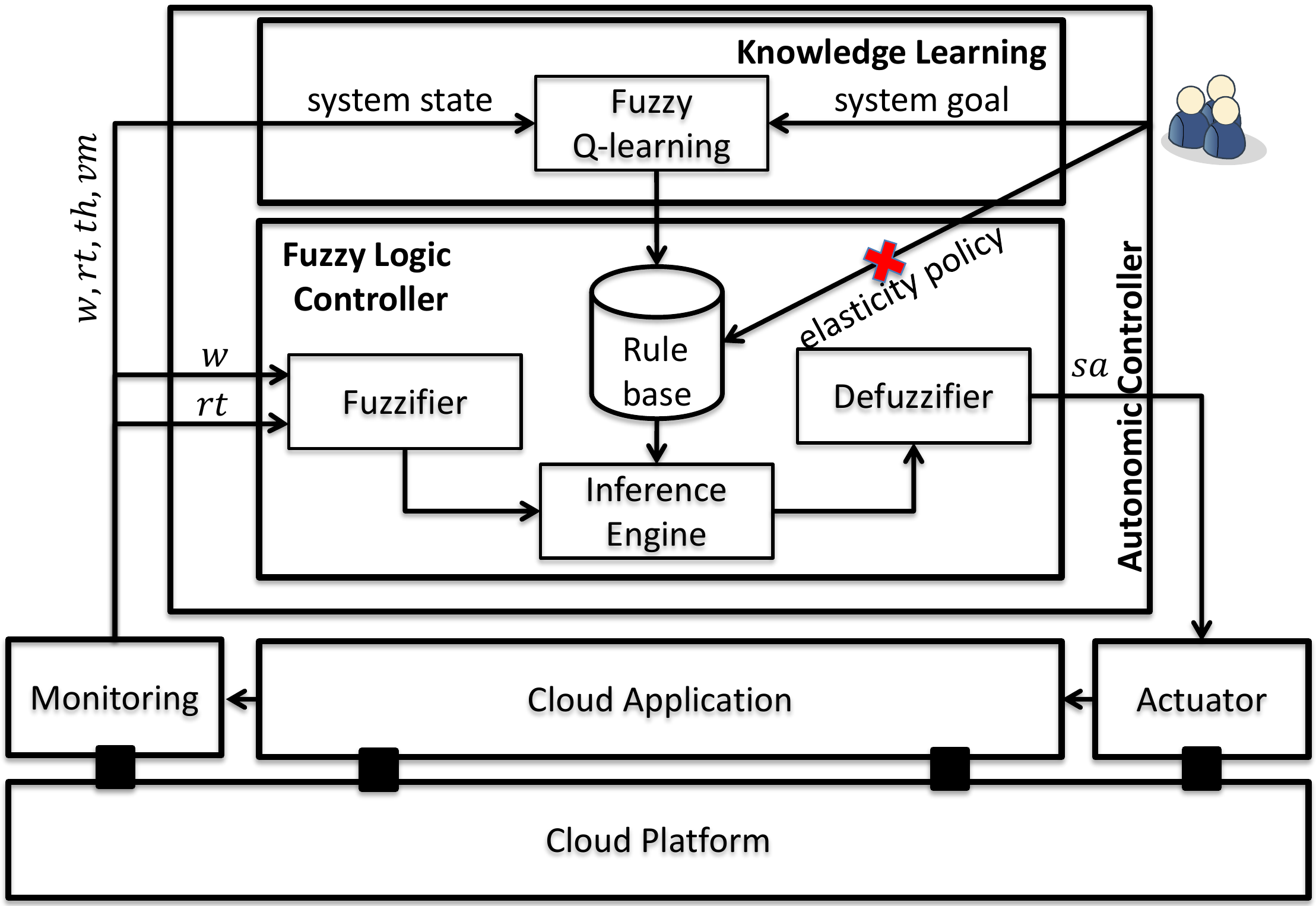}
\caption{FQL4KE architecture.}
\label{fig:fql-architecture}
\end{center}
\end{figure}

Figure \ref{fig:fql-architecture} illustrates the main building blocks of \texttt{FQL4KE}. While the application runs on a cloud platform that provides the demanded resource, \texttt{FQL4KE} monitors the application and guides resource provisioning. More precisely, \texttt{FQL4KE} follows the autonomic MAPE-K loop \cite{Kephart2003}, where different characteristics of the application (e.g. workload and response time) are continuously monitored, the satisfaction of system goals are checked and accordingly the resource allocation is adapted in case of deviation from goals. The goals (i.e., SLA, cost, response time) are reflected in the reward function as we will define this in Section \ref{sec:fqlk}.

The monitoring component collects low-level performance metrics and feed both cloud controller as well as the knowledge learning component. The actuator issues adaptation commands that it receives from the controller at each control interval to the underlying cloud platform. Two components of knowledge learning and cloud controller are incorporated for this purpose. The cloud controller is a fuzzy controller that takes the observed data, and generates scaling actions. The learning component continuously updates the knowledge base of the controller by learning appropriate rules. These two components are described in Sections \ref{sec:flc} and \ref{sec:fql} respectively. Finally, the integration of these two components is discussed in Section \ref{sec:fqlk}.

\subsection{Fuzzy Logic Controller}
\label{sec:flc}

Fuzzy inference is the process of mapping a set of control inputs to a set of control outputs through fuzzy rules. The main application of fuzzy controllers is for types of problems that cannot be represented by explicit mathematical models due to high non-linearity of the system. Instead, the potential of fuzzy logic lies in its capability to approximate that non-linearity by expressing the knowledge in a similar way to the human perception and reasoning \cite{jamshidi2014autonomic}.
The inputs to the controller are the workload (\emph{$w$}) and response time (\emph{$rt$}) and the output is the scaling action (\emph{$sa$}) in terms of increment (or decrement) in the number of virtual machines (VMs).

\begin{figure}
	\centering
	\begin{subfigure}
		\centering
		\includegraphics[width=4.2cm]{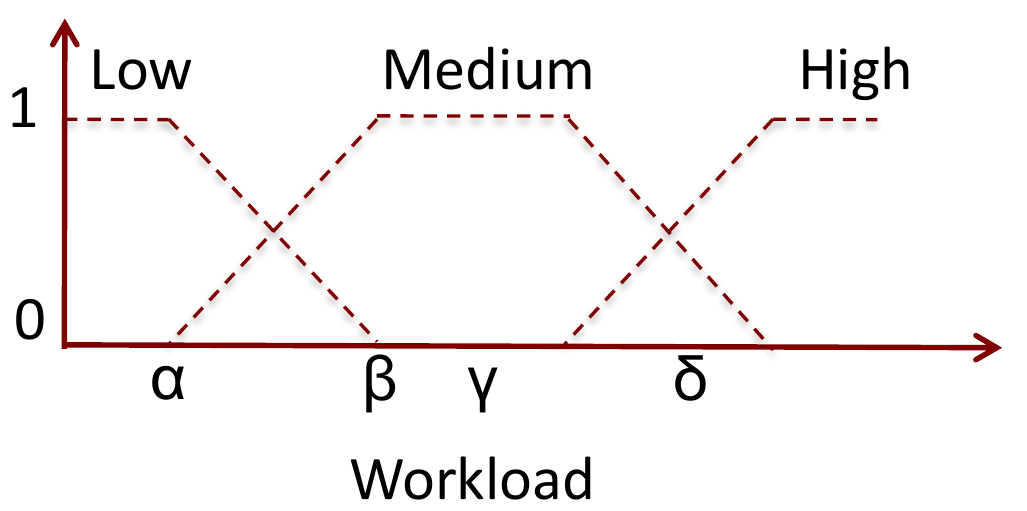}
		\label{fig:workload}
	\end{subfigure}
	\quad
	\begin{subfigure}
		\centering
		\includegraphics[width=4cm]{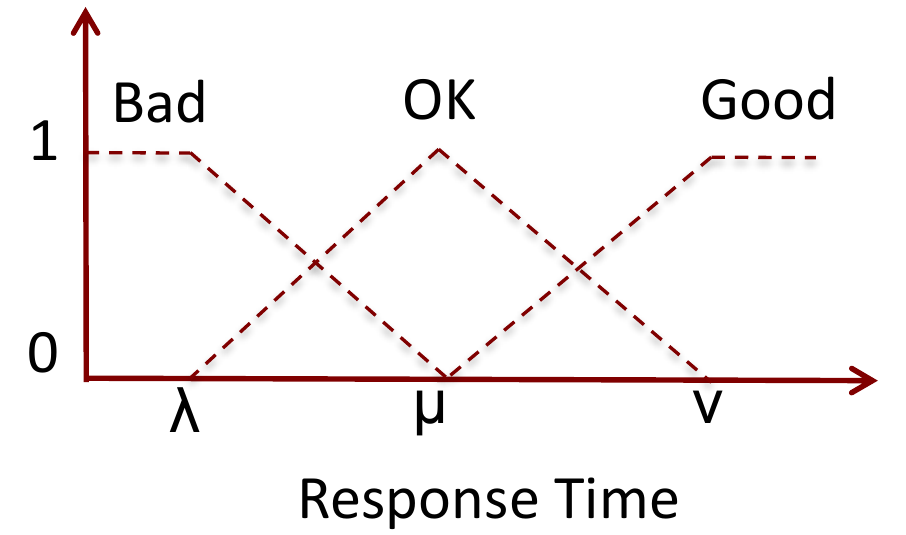}
		\label{fig:responsetime}
	\end{subfigure}
	\caption{Fuzzy metrics for auto-scaling.}
	\label{fig:fuzzymetrics}
\end{figure}

The design of a fuzzy controller, in general, involves the following tasks: 
\begin{inparaenum}
	\item defining the fuzzy sets and membership functions of the input signals.
	\item defining the rule base which determines the behavior of the controller in terms of control actions using the linguistic variables defined in the previous task.
\end{inparaenum} 
The very first step in the design process is to partition the state space of each input variable into various fuzzy sets through membership functions. Each fuzzy set associated with a linguistic term such as ''low'' or ''high''.  The membership function, denoted by $\mu_y(x)$, quantifies the degree of membership of an input signal $x$ to the fuzzy set $y$ (cf. Figure \ref{fig:fuzzymetrics}). In this work, the membership functions, depicted in Figure \ref{fig:fuzzymetrics}., are considered to be both triangular and trapezoidal. As shown, three fuzzy sets have been defined for each input (i.e., workload and response time) to achieve a reasonable granularity in the input space while keeping the number of states small to reduce the set of rules in the knowledge base. This number also corresponds to the number of states we have in Q-learning that we will describe later in the next section. 

The next step consists of defining the inference machinery for the controller by expressing the elasticity policies in linguistic terms as a set of rules. An example for such elasticity policy is: "\texttt{IF} ($w$ is $high$) \texttt{AND} ($rt$ is $bad$) \texttt{THEN} ($sa=+2$)", where the output function is a constant value, which can be an integer in $\{-2,-1,0,+1,+2\}$, which is associated to the change in the number of deployed nodes. Note that this set can be any finite set but here for simplicity we constraint it to only 5 possible actions, but depending on the problem at hand can be any finite discrete set of actions. 
For the definition of the functions in the rule consequents, the knowledge and experience of a human expert is generally used. 
In this work, no a priori knowledge for defining such rules is assumed. In particular, \texttt{FQL4KE} attempts to find the consequent $Y$ for the rules, see Section \ref{sec:fql}. 

Once the fuzzy controller is designed, the execution of the controller is comprised of three steps (cf. middle part of Figure \ref{fig:fql-architecture}): (i) fuzzification of the inputs, (ii) fuzzy reasoning, and (iii) defuzzification of the output. Fuzzifier projects the crisp data onto fuzzy information using membership functions. Fuzzy engine reasons on information based on a set of fuzzy rules and derives fuzzy actions. Defuzzifier reverts the results back to crisp mode and activates an adaptation action. For the sake of simplicity, we calculate the output as a weighted average:

\begin{equation}
\label{eq:weighted-average}
y(x)=\sum_{i=1}^N\mu_i(x) \times a_i,
\end{equation}

where $N$ is the number of rules, $\mu_i(x)$ is the firing degree of the rule $i$ for the input signal $x$ and $a_i$ is the consequent function for the same rule. 
Then the output is rounded to the nearest integer, due to the discrete nature of scaling actions for cloud scaling. Finally, this value, if endorsed by policy enforcer module (see Section \ref{sec:realization}), will be enacted by issuing appropriate commands to the underlying cloud platform fabric.

Fuzzy controller, such as \texttt{RobusT2Scale} that we described in this section, have some limitations. The knowledge base defined to adjust resource to keep the performance of a cloud application at a desired level should be able to be applied to any potential scenario (e.g., different load levels). However, performance of the controller is not always consistent \cite{jamshidi2014autonomic} with the desires of the designer of the controller. The guilty here is the fixed fuzzy rules. In the next, we describe a mechanism to overcome this limitation. 

%
%
%

\subsection{Fuzzy Q-Learning}
\label{sec:fql}

Until this stage, we have shown how to design a fuzzy controller for auto-scaling a cloud-based application where the elasticity policies are provided by users at design-time. In this section, we introduce a mechanism to learn/adjust/adapt the policies at runtime, enabling knowledge evolution (i.e., \texttt{KE} part in \texttt{FQL4KE}) that we promised earlier. 
As the controller has to take an action in each control loop, it should try to select those actions taken in the past which produced good rewards. Here by reward we mean ''long-term cumulative'' reward:

\begin{equation}
\label{eq:cumulative-reward}
R_t=r_{t+1}+\gamma r_{t+2}+\dots=\sum_{k=0}^{\infty} \gamma^k r_{t+k+1},
\end{equation}
where $\gamma$ is the discount rate determining the relative importance of future rewards, in the same way that promising money for some time in the future worth less than the same money for literally now in time. If the agent only takes actions based on the actions that have been already tried, it will stick to a suboptimal knowledge 
Therefore, there exists a trade-off (cf. step 2 in Algorithm \ref{alg:fql}) between the actions that have already tried (known as exploitation) and new actions that may lead to better rewards in the future (known as exploration). 

In each control loop, the controller needs to take an action based on a function of the state in which it is located and the action that it selects. $Q(s,a)$ denotes this Q function based on which the controller takes actions determining the expected cumulative reward that can be received by taking action $a$ in state $s$. This value directly depends on the policy followed by the controller, thus determining the behavior of the controller. This policy $\pi(s,a)$ is the probability of taking action $a$ from state $s$. As a result, the value of taking action $a$ in state $s$ following the policy $\pi$ is formally defined as:
\begin{equation}
\label{eq:qpi}
Q^{\pi}(s,a)=E_{\pi}\{\sum_{k=0}^{\infty}\gamma^kr_{t+k+1}\},
\end{equation}
where $E_{\pi}\{.\}$ is the expectation function under policy $\pi$. When an appropriate policy is found, the RL problem at hand is solved. Q-learning is a technique that does not require any specific policy in order to evaluate $Q(s,a)$:
\begin{equation}
\label{eq:q-update}
Q(s_t,a_t) \leftarrow Q(s_t,a_t)+\eta [r_{t+1}+\gamma \max_aQ(s_{t+1},a)-Q(s_t,a_t)],
\end{equation}
where $\gamma$ is the learning rate. Therefore, the optimal Q function can be approximated without any specific policy to follow. In this case, the policy adaptation can be achieved by selecting a random action with probability $\epsilon$ and an action that maximizes the Q function in the current state with probability $1-\epsilon$, note that the value of $\epsilon$ is determined by the exploitation/exploration strategy (cf. \ref{sec:setting}):
\begin{equation}
\label{eq:greedy-policy}
a(s)=arg \max_k Q(s,k)
\end{equation}

Fuzzy logic version of Q-learning \cite{Glorennec1997} optimizes the consequents of the rules in fuzzy controllers. Fuzzy Q-learning (FQL) has some critical benefits over its traditional algorithm. First and most importantly, for some application areas in which the number of states and the potential actions that the agent can take in those states are high; hence the q-values need to be stored in large look up tables. As a result, the Q-learning becomes impractical in continuous state spaces \cite{Glorennec1997} such as our case in this paper. By employing fuzzy variables, continuous state spaces can be discretized into states represented by all the combinations of variables (cf. Figure \ref{fig:fuzzymetrics}). 

\begin{tiny} 
	\begin{algorithm}
		\caption{: Fuzzy Q-Learning}
		\label{alg:fql}
		\begin{algorithmic}[1]
			\Require $\gamma, \eta $			
			\State \text{Initialize q-values:}
			\break $q[i,j]=0,~1<i<N~,~1<j<J$
			\State \text{Select an action for each fired rule:} 
			\break $a_i=arg max_kq[i,k]$ with probability $1-\epsilon$ \Comment{\textbf{Eq. \ref{eq:greedy-policy}}}
			\break $a_i=random\{a_k, k=1,2,\cdots,J\}$ with probability $\epsilon$
			\State \text{Calculate the control action by the fuzzy controller:} 
			\break $a=\sum_{i=1}^N\mu_i(x) \times a_i,$ \Comment{\textbf{Eq. \ref{eq:weighted-average}}}
			\break where $\alpha_i(s)$ is the firing level of the rule $i$
			\State Approximate the Q function from the current q-values and the firing level of the rules:
			\break $Q(s(t),a)=\sum_{i=1}^N\alpha_i(s)\times q[i,a_i],$
			\break where $Q(s(t),a)$ is the value of the Q function for the state current state $s(t)$ in iteration $t$ and the action $a$ 
			\State Take action $a$ and let system goes to the next state $s(t+1)$.
			\State Observe the reinforcement signal, $r(t+1)$ and compute the value for the new state:
			\break $V(s(t+1))=\sum_{i=1}^N\alpha_i(s(t+1)).max_k(q[i,q_k])$.
			\State \text{Calculate the error signal:}
			\break $\Delta Q=r(t+1)+\gamma \times V_t(s(t+1))-Q(s(t),a),$ \Comment{\textbf{Eq. \ref{eq:q-update}}}
			\break where $\gamma$ is a discount factor
			\State \text{Update q-values:}
			\break $q[i,a_i]=q[i,a_i]+\eta \cdot \Delta Q \cdot \alpha_i(s(t)),$ \Comment{\textbf{Eq. \ref{eq:q-update}}}
			\break where $\eta$ is a learning rate
			\State Repeat the process for the new state until it converges
		\end{algorithmic}
		\vspace{2mm}
	\end{algorithm}
\end{tiny}

The fuzzy Q-learning algorithm that we have implemented is summarized in Algorithm \ref{alg:fql}. In the case of our running example, the state space is finite (i.e., 9 states as the full combination of $3 \times 3$ membership functions for fuzzy variables $w$ and $rt$) and \texttt{RobusT2Scale} has to choose a scaling action among 5 possible actions $\{-2,-1,0,+1,+2\}$. However, the design methodology that we demonstrated in this section is general and can be applied for any possible state and action spaces. Note that the convergence is detected when the change in the consequent functions is negligible in each learning loop. 

\subsection {FQL4KE for Dynamic Resource Allocation}
\label{sec:fqlk}

The combination of the fuzzy controller and the fuzzy Q-learning algorithm is illustrated in Figure \ref{fig:fql-architecture}. 

\emph{Reward function.} As illustrated in Figure \ref{fig:fql-architecture}, the controller receives the current values of $w$ and $rt$ that correspond to the state of the system, $s(t)$ (cf. Step 4 in Algorithm \ref{alg:fql}). The control signal $sa$ represents the action $a$ that the controller needs to take at each state. We define the reward signal $r(t)$ based on three criteria: (i) SLO violations, (ii) the amount of resource acquired, which directly determine the cost, and (iii) throughput, as follows:

\begin{equation}
\label{eq:reinforcement-signal}
r(t)=U(t)-U(t-1),
\end{equation}
where $U(t)$ is the utility value of the system at time $t$. Hence, if a controlling action leads to an increased utility, it means that the action is appropriate. Otherwise, if the reward is close to zero, it implies that the action is not effective. A negative reward (punishment) warns that the situation is worse after taking the action. The utility function that we defined is as below:

\begin{equation}
\label{eq:utility-function}
U(t) = w_1 \cdot {th(t) \over th_{max}} + w_2 \cdot (1-{vm(t) \over vm_{max}})+w_3 \cdot (1-H(t))
\end{equation}
\[
H(t)= 
\begin{cases}
{(rt(t)-rt_{des}) \over rt_{des}} & rt_{des} \leq rt(t)\leq 2\cdot rt_{des}\\
1 & rt(t) \geq 2\cdot rt_{des} \\
0 & rt(t) \leq rt_{des} 
\end{cases}
\]
where $th(t)$, $vm(t)$ and $rt(t)$ are throughput, number of worker roles (VMs) and response time of the system, respectively. $w_1$,$w_2$ and $w_3$ are their corresponding weights determining their relative importance in the utility function. Note that in order to aggregate the individual criteria together, we normalized them depending on whether they should be maximized or minimized. It is important to know that the only input which is required by the users to determine the system goals is the value for these three weights, i.e., $w_1$,$w_2$ and $w_3$ (cf. Figure \ref{fig:fql-architecture}). 

\emph{Knowledge base update.} \texttt{FQL4KE} starts with controlling the allocation of resources with no a priori knowledge. After enough \emph{explorations}, the consequents of the rules can be determined by selecting those actions that correspond to the \emph{highest} q-value in each row of the Q table. Although \texttt{FQL4KE} does not rely on design-time knowledge, if even partial knowledge is available (i.e., operator of the system is confident with providing some of the elasticity policies) or there exists data regarding performance of the application, \texttt{FQL4KE} can exploit such knowledge by initializing q-values (cf. step 1 in Algorithm \ref{alg:fql}) with more meaningful data instead of initializing them with zero. This implies a quicker learning convergence.

\section{Realization}
\label{sec:realization}

\begin{figure}
	\begin{center}
		\includegraphics[width=9cm]{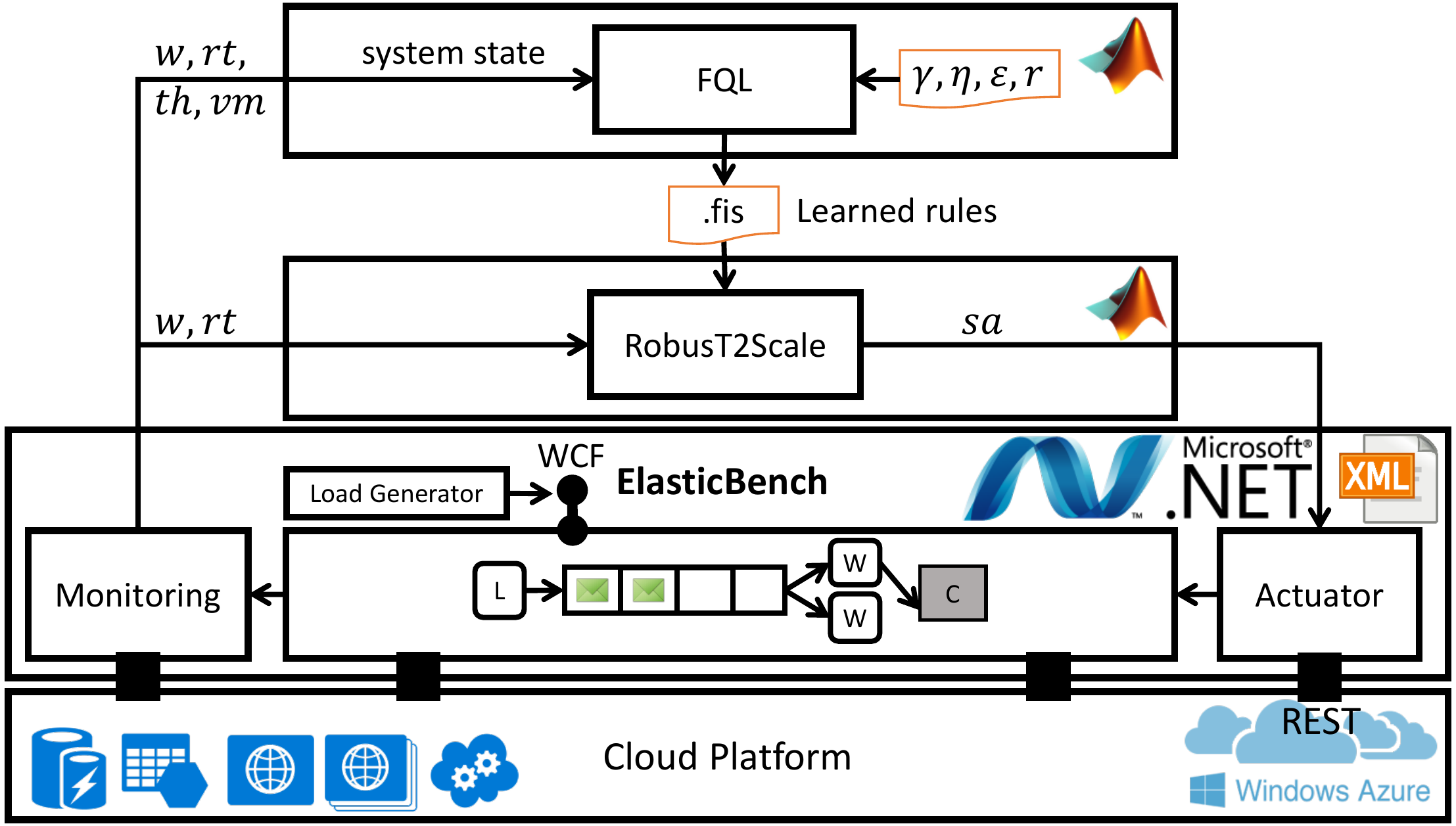}
		\caption{FQL4KE realization}
		\label{fig:implementation}
	\end{center}
\end{figure}

To demonstrate the applicability of our approach, we realized a prototype of \texttt{FQL4KE} as a generic, configurable and customizable platform on Microsoft Azure. As illustrated in Figure \ref{fig:implementation}, this prototype comprises of 3 integrated components:
\begin{enumerate}[i]
	\item A learning component \texttt{FQL} implemented in Matlab \footnote{code is available at https://github.com/pooyanjamshidi/Fuzzy-Q-Learning}
	\item A cloud controller reasoning engine (\texttt{RobusT2Scale}) implemented in Matlab \footnote{code is available at https://github.com/pooyanjamshidi/RobusT2Scale},
	\item A cloud-based application framework (\texttt{ElasticBench}) implemented with Microsoft .NET technologies (.NET framework 4 and Azure SDK 2.5) \footnote{code is available at https://github.com/pooyanjamshidi/ElasticBench}
	\item The integration between these three components by software connectors developed with .NET technologies.
\end{enumerate}

The rule base of \texttt{FQL4KE} is continuously updated by \texttt{FQL} component and new/updated rules are fed into the cloud controller \texttt{RobusT2Scale} as an intermediate text-based \emph{*.fis} file. \texttt{ElasticBench} is a new component we developed with Microsoft development technologies and sits on top of the cloud platform. \texttt{ElasticBench} is in charge of monitoring the application as well as deploying virtual machines. 

\subsection{ElasticBench}

\begin{figure}
	\begin{center}
		\includegraphics[width=9cm]{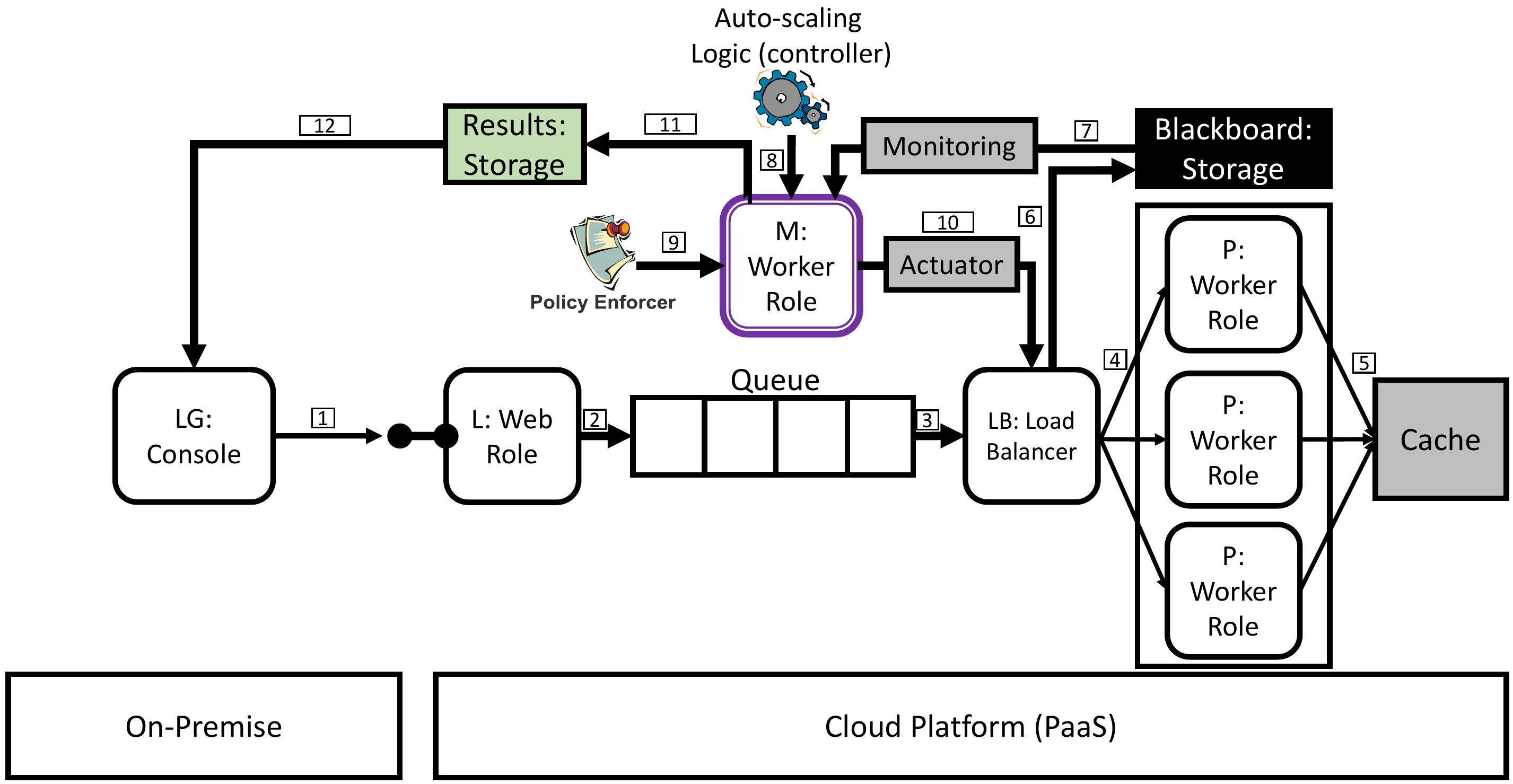}
		\caption{Overview of our experimental setting.}
		\label{fig:platform}
	\end{center}
\end{figure}

ElasticBench includes a workload generator in order to simulate different patterns. This allows to test and train the controller before actual execution. It also provides all the required functionalities to perform a variety of auto-scaling experiments. In order to build a generic workload generator, we developed a service to generate Fibonacci numbers. A delay is embedded in the process of calculating Fibonacci numbers to simulate a process that takes a reasonably long period. Note that calculating Fibonacci numbers is an $O(N)$ task, which makes it a good candidate for demonstrating different application types by embedding different delays, since our platform can generate requests with varying patterns. As depicted in Figure \ref{fig:platform} a positive integer ($n \in N$) is written to the queue and on the other end of the queue, a Fibonacci series should be calculated, and this takes different times to be calculated depending on the number of processor at the other end of the queue. This enables us to test our solution with a generic platform that has several functionalities, each of which takes different times depending on the available resources providing different level of processing capabilities. This resemble a real-world software system that expose different functionalities with different response time depending on the resource contention. For instance, some functionalities are CPU intensive and depending on the number of available CPU reply with a different response time, while others may be memory intensive and depending on the available memory takes different time to react.

Two types of Azure services are used to implement ElasticBench: \emph{web role} and \emph{worker role}. Note that web and worker roles corresponds to VM at infrastructure level. The requests issued from load generator are received by the web role, which puts a message on a task assignment queue as shown in Figure \ref{fig:platform}. 
The worker role instances continuously checks this queue and after locating a message, a background process (to calculate Fibonacci number) is immediately started based on the content of the message in the queue. The worker roles communicate with the storage to acquire the data required for processing (e.g., previously calculated Fibonacci numbers).

We implemented two types of worker role: some worker roles (\emph{P}) process the messages (i.e., calculating the Fibonacci numbers), whereas the other type of worker role (\emph{M}) implements the MAPE-K feedback control loop. 
The main functionalities in \emph{M} worker role is as follows: (1) It reads performance metrics from the blackboard storage; (2) It calculates metrics for feeding the fuzzy controller; (3) It also implements a policy enforcer to check whether the number of nodes to be enacted is within the predefined range and whether the worker role is in a stable mode. (4) It is possible to plug-in other cloud controllers (i.e., controllers implementing other techniques) with few lines of code; (5) It also implements mechanisms comprising the resiliency of this worker role.

The design decision we made for implementing the MAPE-K functionalities inside a worker role in the cloud was strategic for the experiments that we needed to run. In one hand, in order to avoid network latencies for decision enaction, we required an internal and isolated network between the decision maker module (i.e., \emph{M}) and the scaling roles (i.e., \emph{P}). On the other hand, we needed to provide a close design to the real world setting as it is the case for commercial solutions in public clouds that the auto-scaling controller sits near the scaling layer as opposed to be deployed on premise. 

In summary, ElasticBench follows these steps in each experiment: (1) workload generator \emph{LG} generates a set of requests according to a predefined pattern. (2) The listener service \emph{L} receives the requests and push them as messages to task assignment queue. (3,4,5) Worker roles \emph {P} picks the messages from the queue and process them. If the results are stored in the cache beforehand, they only read them from the cache. (6) Low-level monitoring performance metrics are stored in the blackboard storage. (7) Worker role \emph{M} retrieves the counters and calculates the metrics. (8) \emph{M} then feed them to the cloud controller and a decision (i.e., change in the number of \emph{P} worker roles) is taken. (9) If the decision has been endorsed by the policy enforcer \footnote{Policy enforcer checks the decision against user-defined constraints} then (10) it is enacted by the actuator to the underlying platform. Note that the actuator calls the appropriate RESTFull operations of the Azure platform in order to change the configuration of the \emph{P} worker roles and to enact the changes accordingly (11) the worker role \emph{M} periodically writes the results to a table storage, (12) which can later be used for further analysis.

%
%

\subsection{Online Knowledge Learning}
\label{sec:online-learning-realization}

\begin{figure}
	\begin{center}
		\includegraphics[width=9cm]{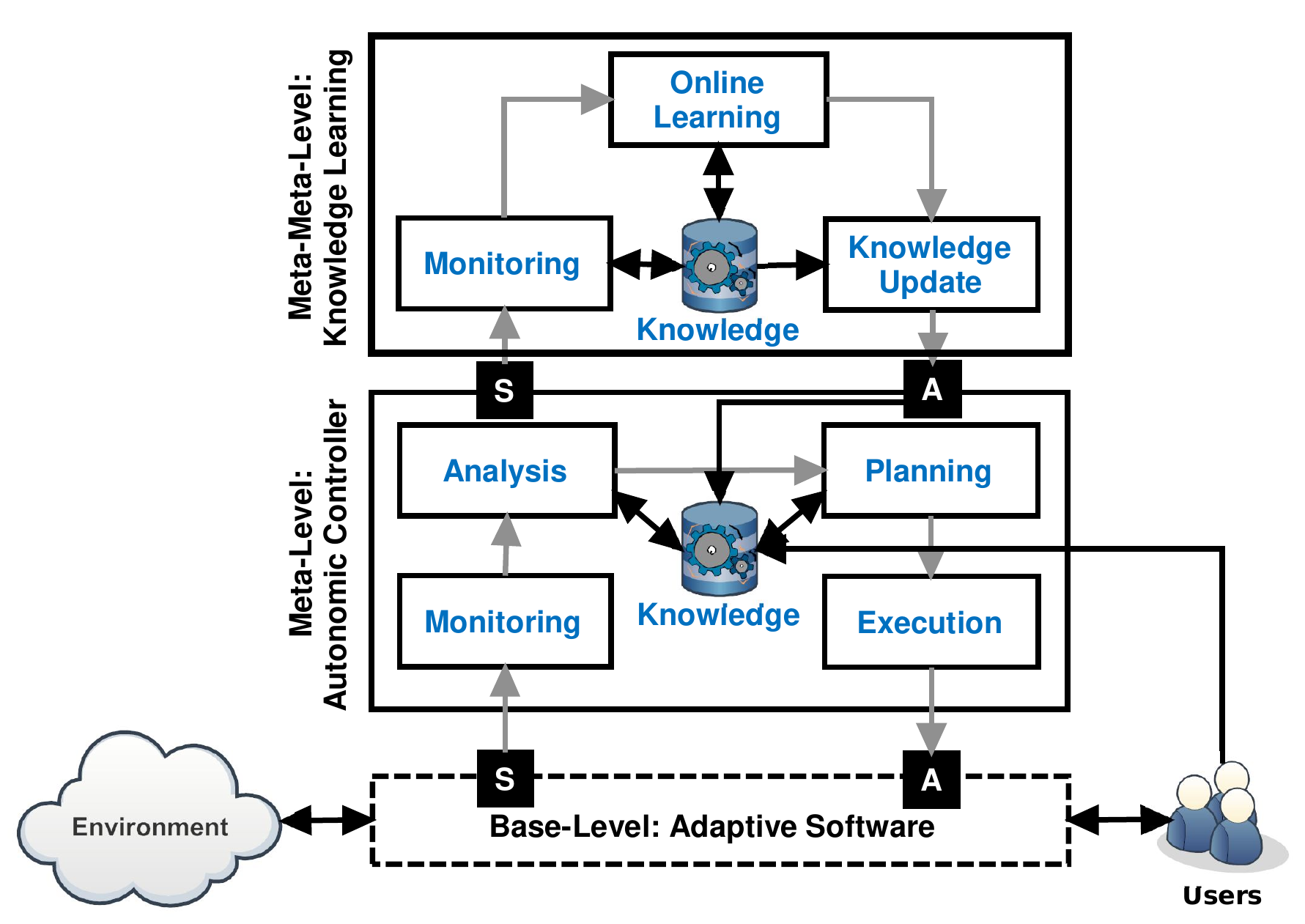}
		\caption{Augmenting MAPE-K with online learning.}
		\label{fig:knowledge-evolution}
	\end{center}
\end{figure}

Online learning is a mechanism for enabling knowledge evolution at runtime \cite{Abbas2011}. As shown in Figure \ref{fig:knowledge-evolution}, online knowledge learning operates on top of the autonomic controller. 
The realization of the learning mechanism is divided in the following phases: (i) monitored operation, (ii) learning, (iii) normal operation. Each phase corresponds to an execution \emph{mode}. The learning process is executed and the system enters to the \emph{monitored operation} mode. In this mode, statistics for analysis is periodically collected. After completion, control is returned to the learning process which enters the \emph{learning} mode. In this phase, depending on the collected analytics, it may update the knowledge base. This completes the knowledge learning loop (cf. loop at the top layer in Figure \ref{fig:knowledge-evolution}). 

%

\section{Experimental Evaluation}
\label{sec:experiment}

We demonstrate the  \emph{efficiency} and \emph{effectiveness} of \texttt{FQL4KE} via an experimental evaluation. More specifically, the key purpose of the experiments is to answer the following questions:

\textbf{RQ1.} Is \texttt{FQL4KE} able to learn how to efficiently acquire resources for dynamic systems in cloud environment?

\textbf{RQ2.} Is \texttt{FQL4KE} flexible enough to allow the operator to set different strategies? and how the approach is effective in terms of key elasticity criteria (cf. criteria column in Table \ref{tab:comparison})?  

\subsection{Experimental Setting}
\label{sec:setting}
\begin{figure}
	\begin{center}
		\includegraphics[width=6cm]{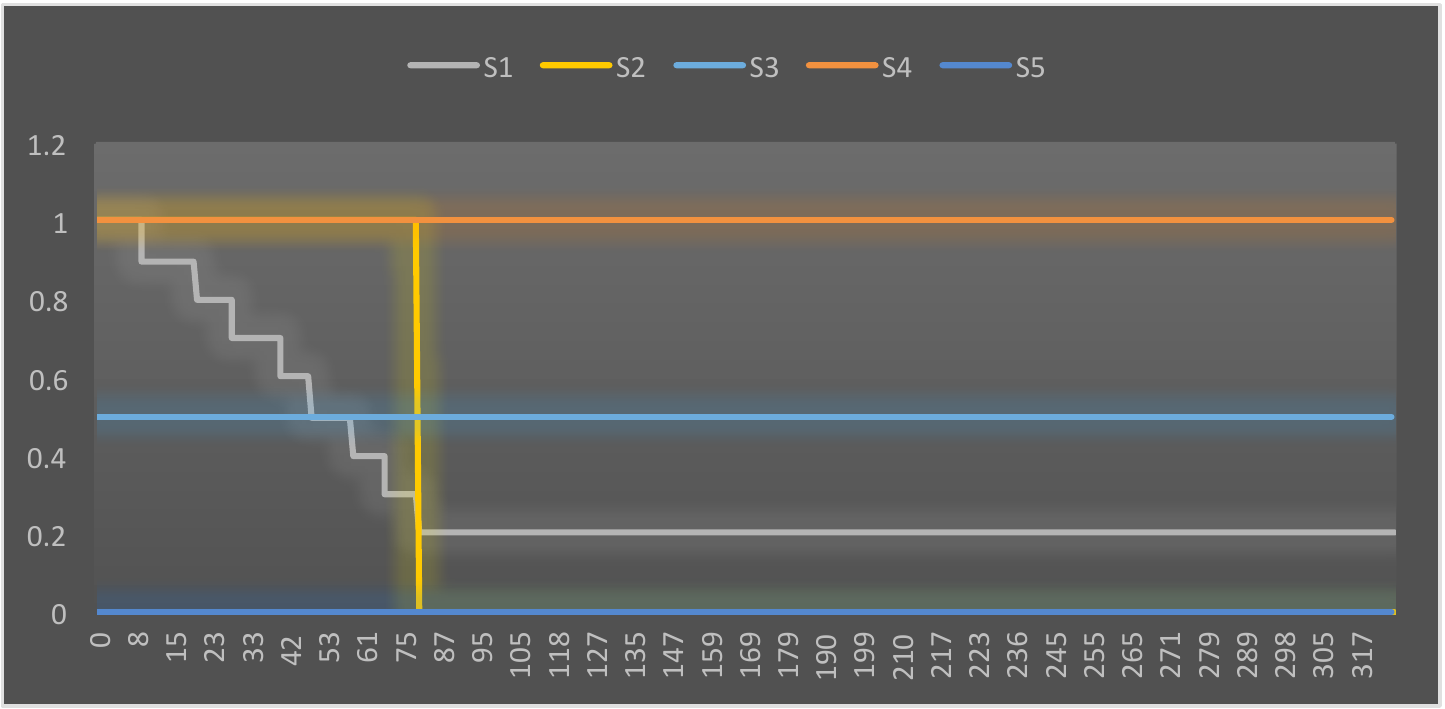}
		\caption{Exploitation/exploration strategies.}
		\label{fig:epsilons}
	\end{center}
\end{figure}

The environment (i.e., cloud platform) in this setting is unique in some aspects. The main differentiating aspects is the delay in receiving rewards after each scaling action has been taken. The agent (i.e., cloud controller) deployed in a delayed-feedback environment (i.e., cloud) comes to know the reward after a non-negative integer indicating the number of time-steps between an agent taking an scaling action and actually receiving its feedback (the state observation and reward). In each monitoring cycle, which happens every 10 seconds, the controller knows about its state but in order to receive the reinforcement signal, it has to wait for example for 8-9 minutes for "scaling out" actions and 2-3 minutes for "scaling in" actions to be enacted. Such kinds of delayed feedback environments introduce some challenges for learning convergence. We tackled this by investigating different learning strategies. As depicted in Figure \ref{fig:epsilons}, we considered 5 different exploitation/exploration strategies (i.e., $S1-S5$). For instance, in $S1$, the learning process starts by a high exploration rate, i.e, $\epsilon=1$ (cf. Step 2 in Algorithm \ref{alg:fql}). We set this in order to explore all possible actions enough times in early cycles. Once the optimal fuzzy rules are learned, the controller with updated elasticity policies will replace the current one. However, at this stage, due to the dynamics of the workload, we cannot set $\epsilon=0$ because a change introduced to the workload, the learning process needs to be performed. As a result, after first set of rules has been learned, we set $\epsilon=0.2$ in order to maintain a balance between exploration and exploitation. In summary, FQL starts with exploration phase and after a first learning convergence happened, it enters the balanced exploration-exploitation phase. However, in order to compare the performance of \texttt{FQL4KE} under different strategies, we consider other learning strategies as well. For instance, in $S2$, after initial learning by high exploration, we set $\epsilon=0$ in order to fully exploit the learned knowledge. 

   
The learning rate in the experiments are set to a constant value $\eta=0.1$ and the discount factor is set to $\gamma=0.8$. The minimum and maximum number of nodes that are allowed to be enacted is set to $1$ and $7$ respectively. We set the control interval to $10 sec$. The worker role that our \texttt{FQL4KE} is deployed is \textit{small} VM with $1$ core and $1792MB$ memory while the \textit{P} worker roles (cf. Figure \ref{fig:platform}) are \textit{extra small} VMs with $1$ core and $768 MB$ memory. Initially, we set all cells in Q table to zero, assuming no a priori knowledge. We set the weights in the reward function all equal, i.e., $w_1=w_2=w_3=1$ (cf. Eq. \ref{eq:utility-function}). The experiment time has been set to $24 hours$ in order to monitor the performance of the system in enough learning steps (on average due to the delay in reward observation, each step takes between $2-9mins$).

\subsection{FQL4KE Efficiency (RQ1)}
\label{sec:rq1}

\begin{figure}
	\centering
	\begin{subfigure} 
		\centering
		\includegraphics[width=2.8cm]{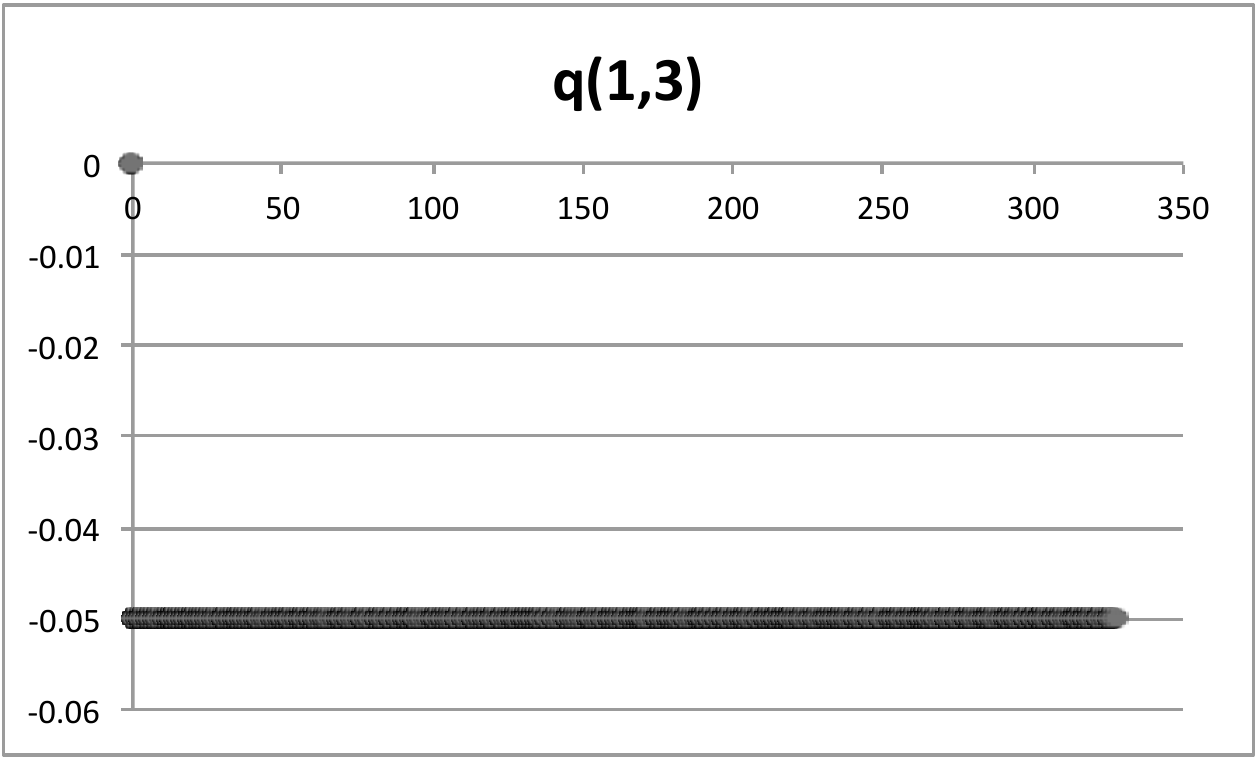}
	\end{subfigure}
	\begin{subfigure} 
		\centering
		\includegraphics[width=2.8cm]{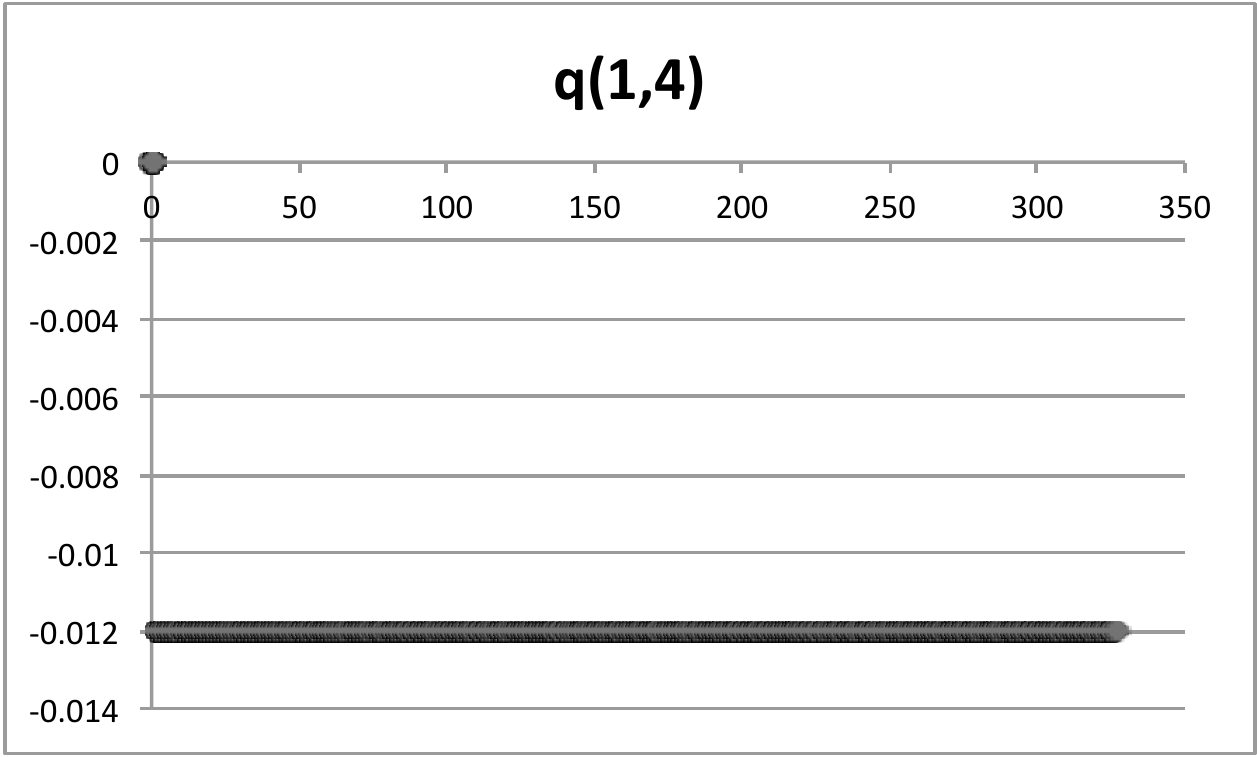}
	\end{subfigure}
	\begin{subfigure} 
		\centering
		\includegraphics[width=2.7cm]{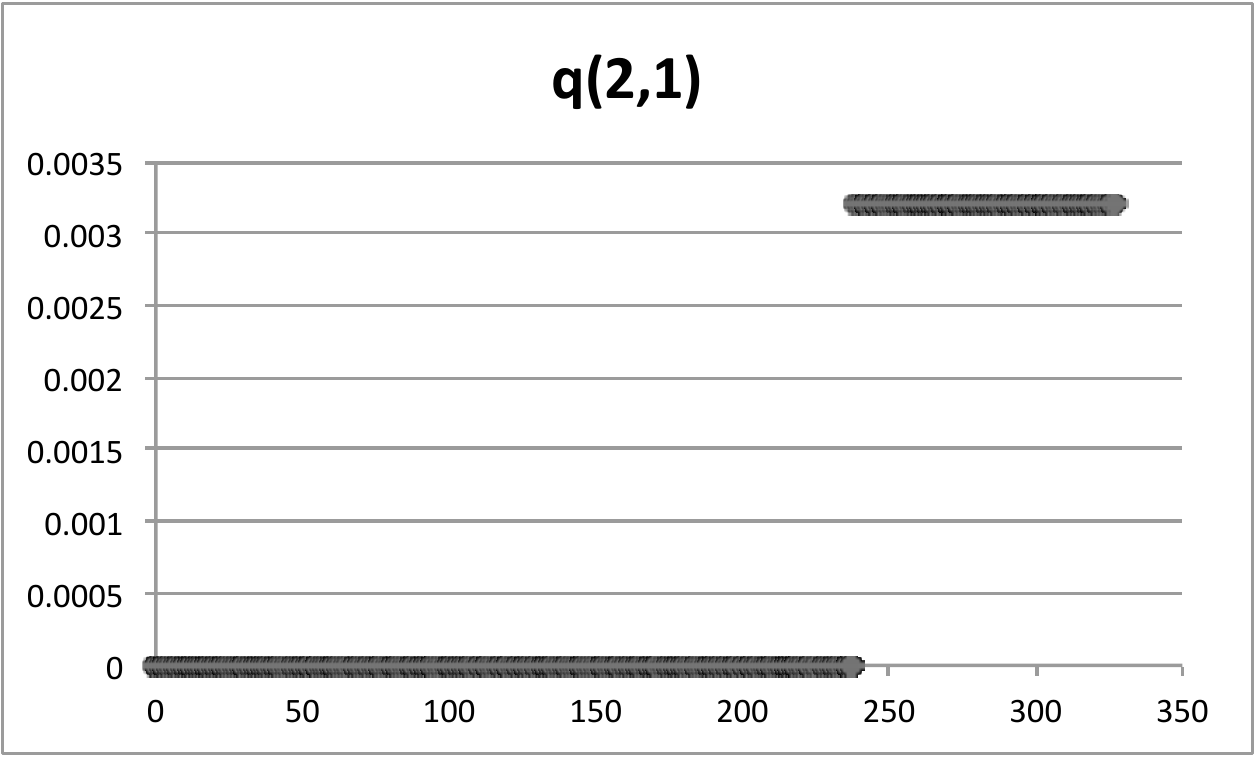}
	\end{subfigure}
	\vfill
	\begin{subfigure} 
		\centering
		\includegraphics[width=2.8cm]{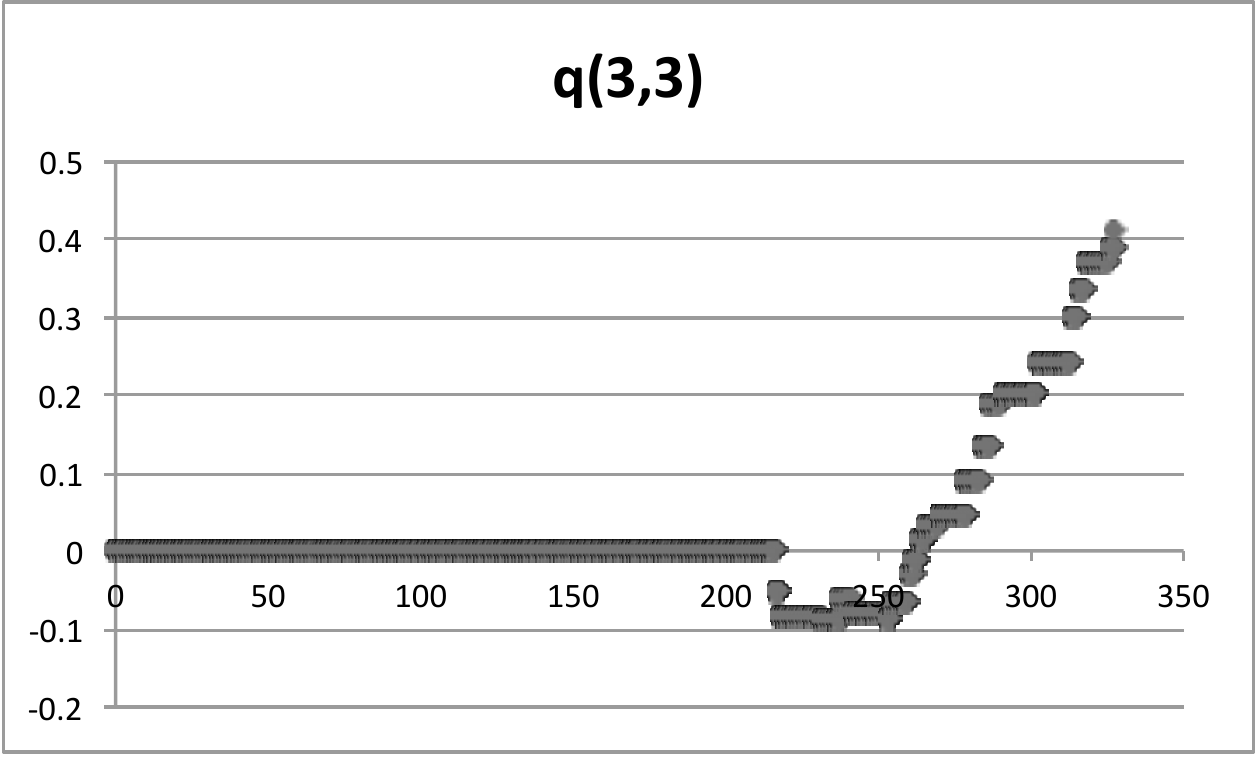}
	\end{subfigure}
	\begin{subfigure} 
		\centering
		\includegraphics[width=2.8cm]{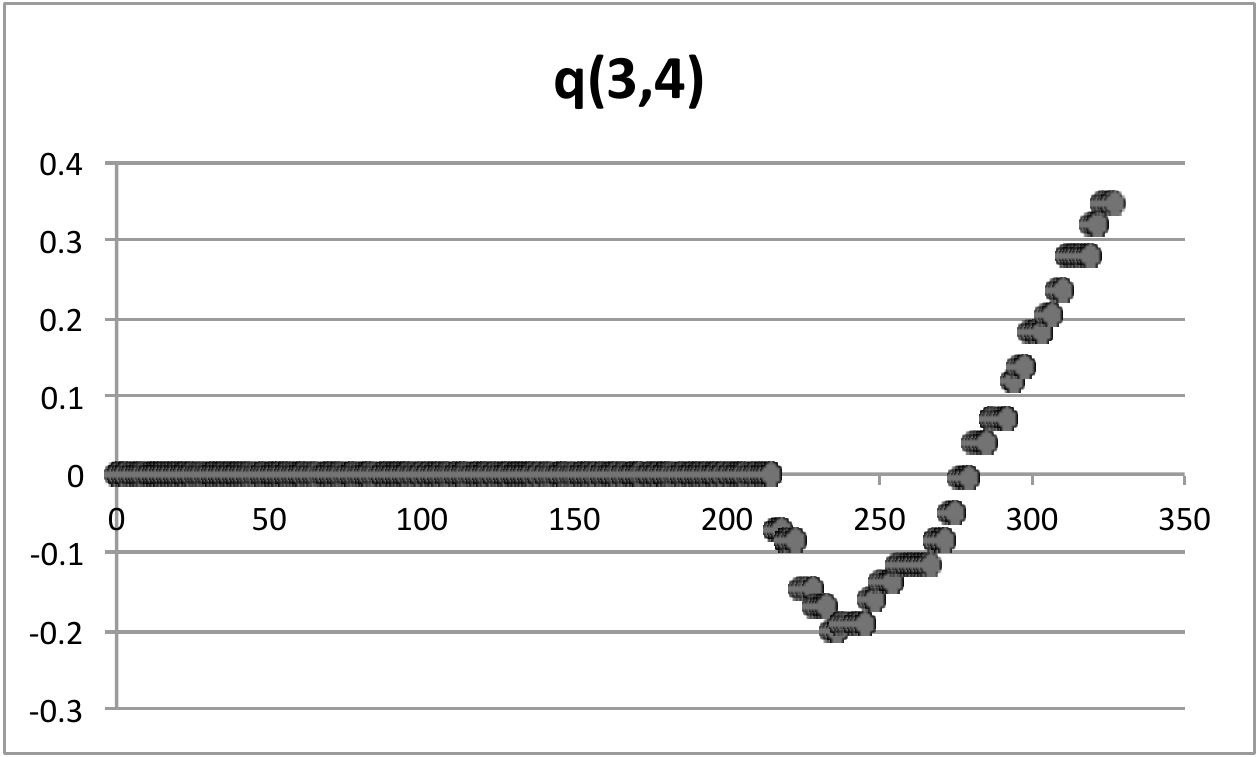}
	\end{subfigure}
	\begin{subfigure} 
		\centering
		\includegraphics[width=2.7cm]{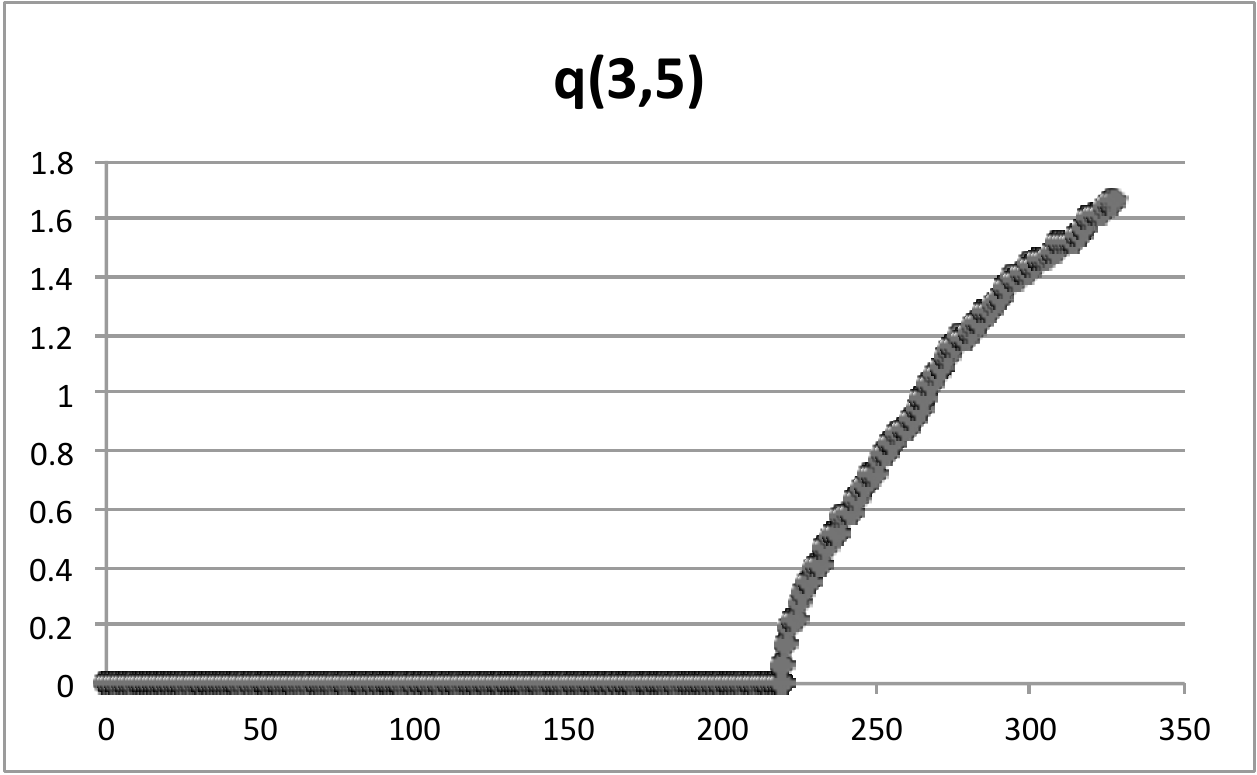}
	\end{subfigure}
	\vfill
	\begin{subfigure} 
		\centering
		\includegraphics[width=2.8cm]{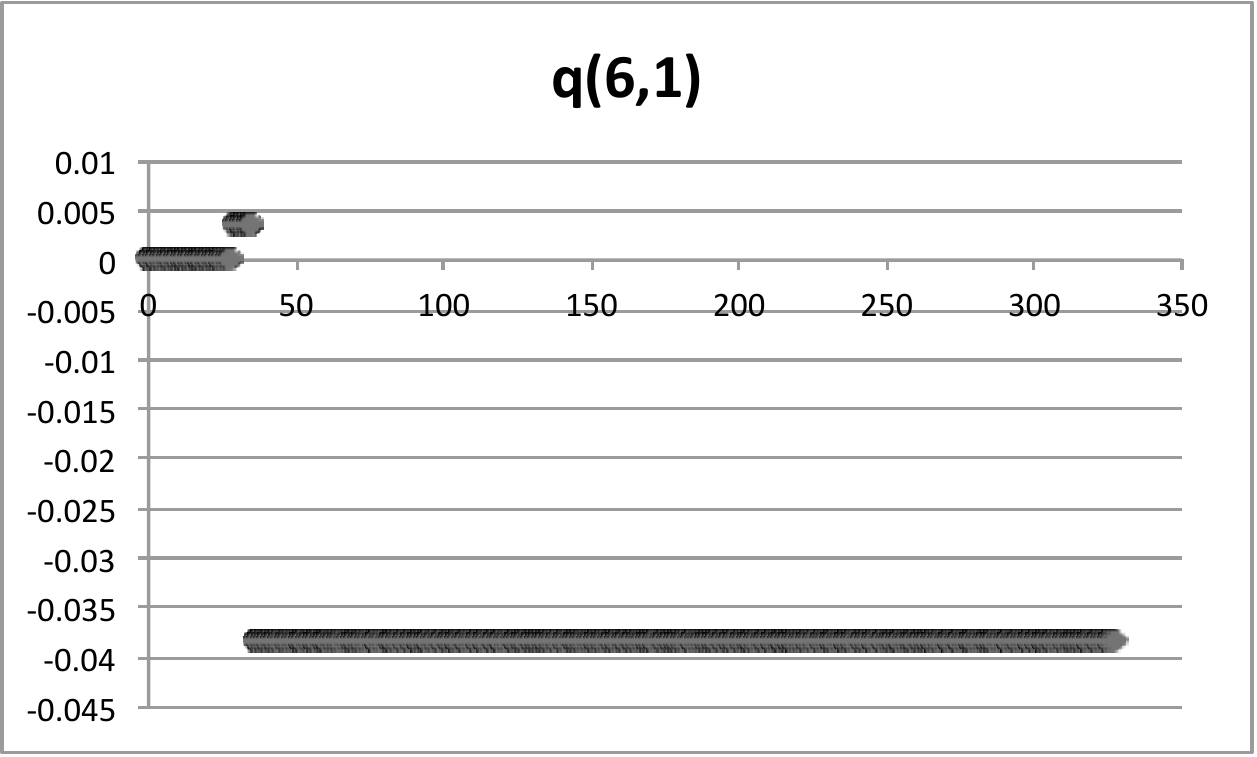}
	\end{subfigure}
	\begin{subfigure} 
		\centering
		\includegraphics[width=2.8cm]{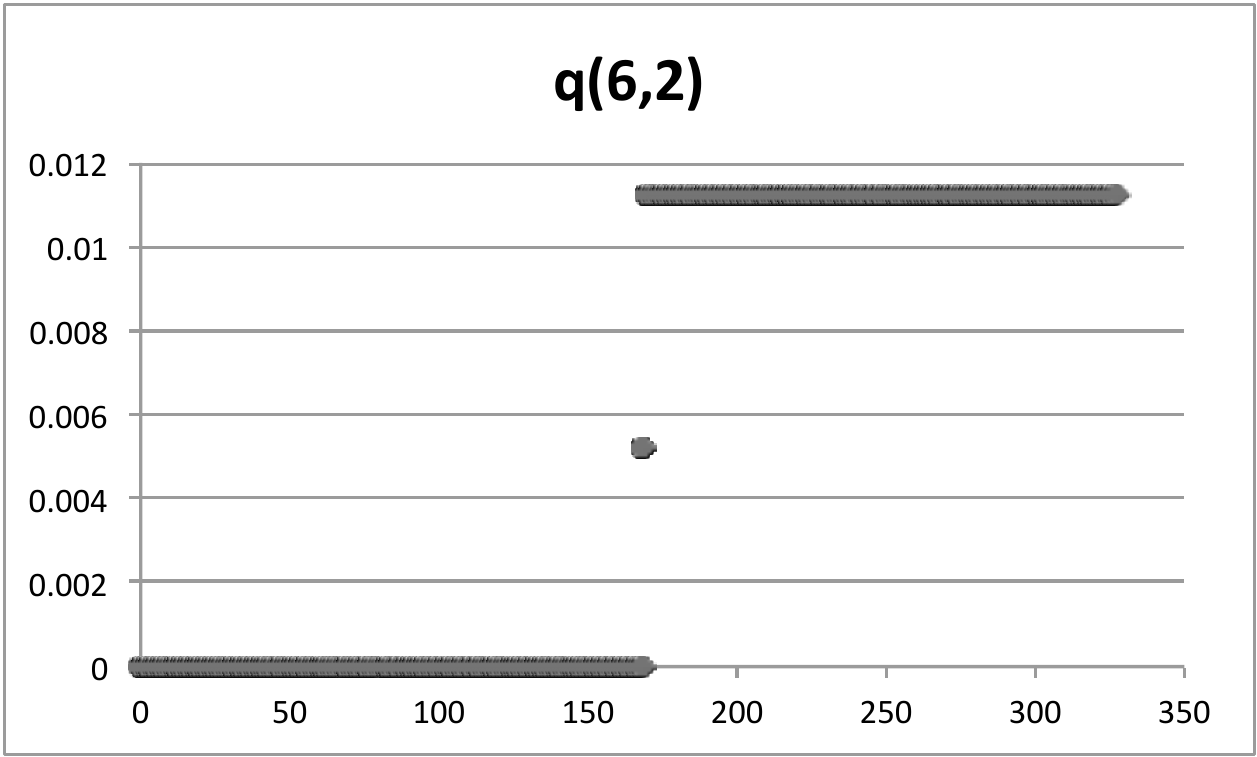}
	\end{subfigure}
	\begin{subfigure} 
		\centering
		\includegraphics[width=2.7cm]{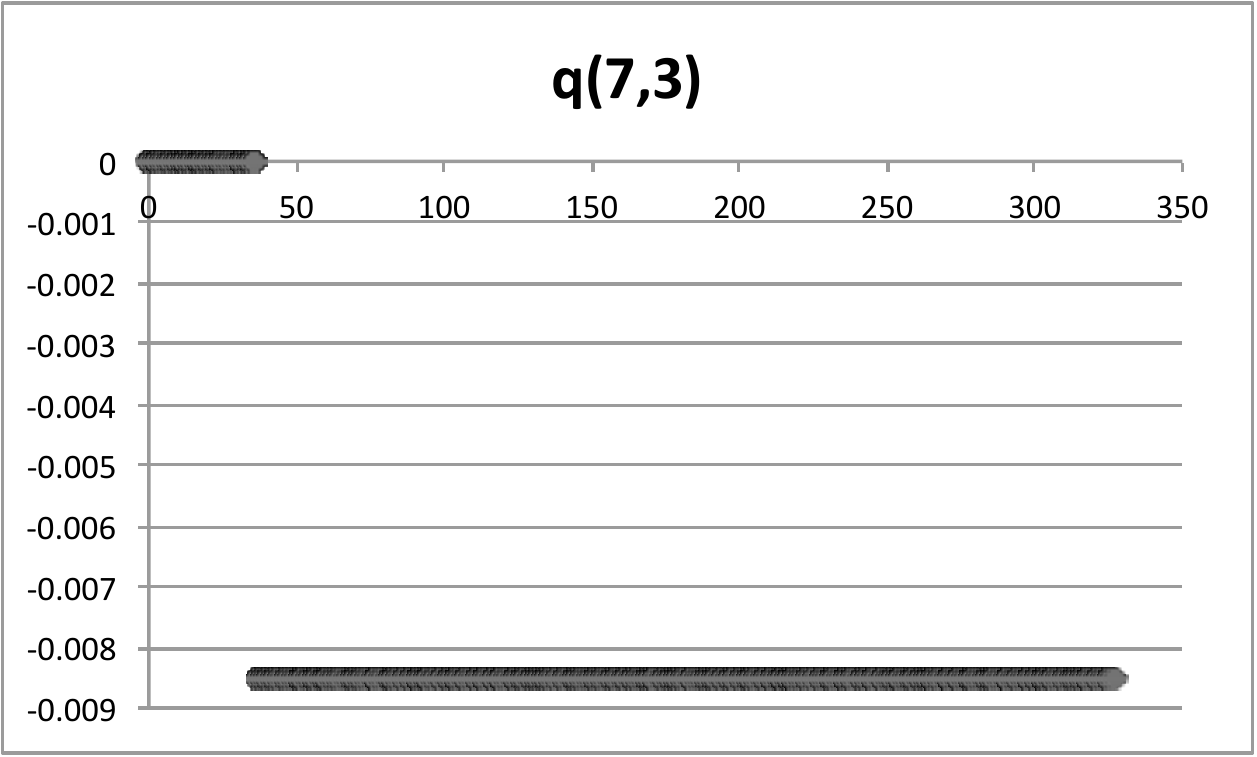}
	\end{subfigure}
	\vfill
	\begin{subfigure} 
		\centering
		\includegraphics[width=2.8cm]{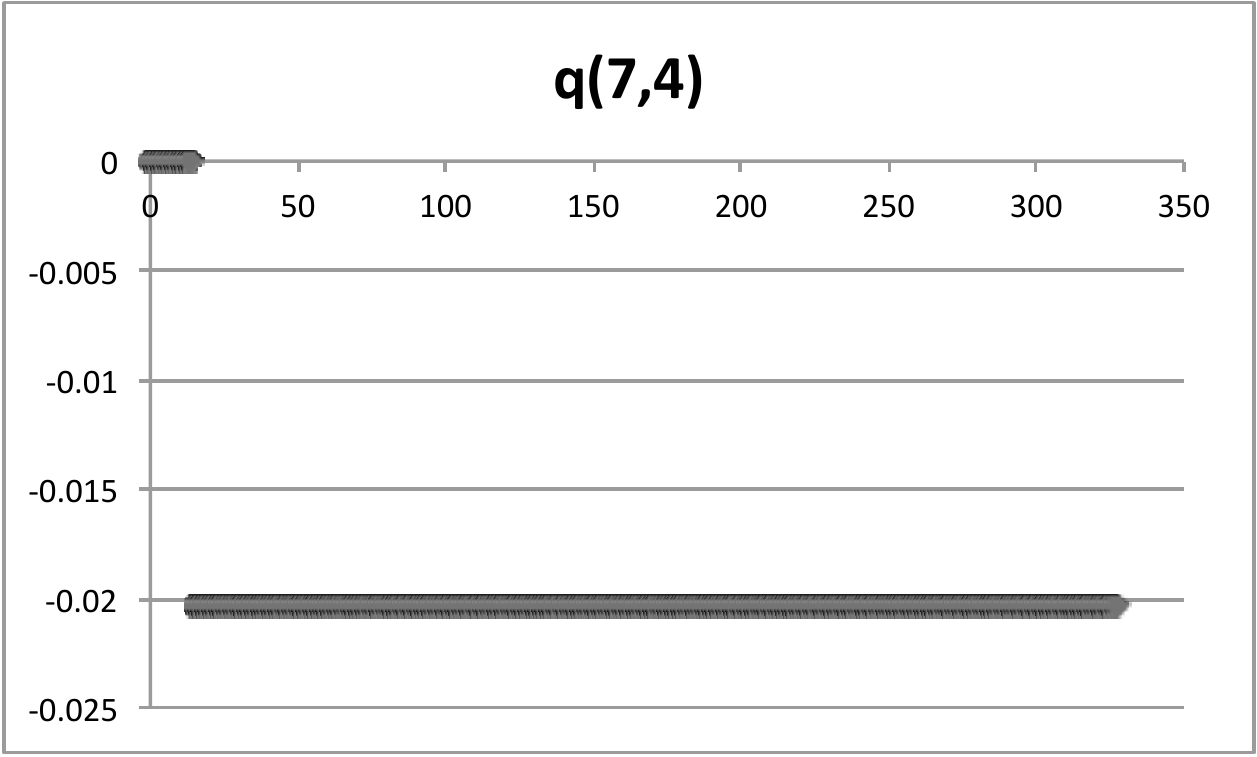}
	\end{subfigure}
	\begin{subfigure} 
		\centering
		\includegraphics[width=2.8cm]{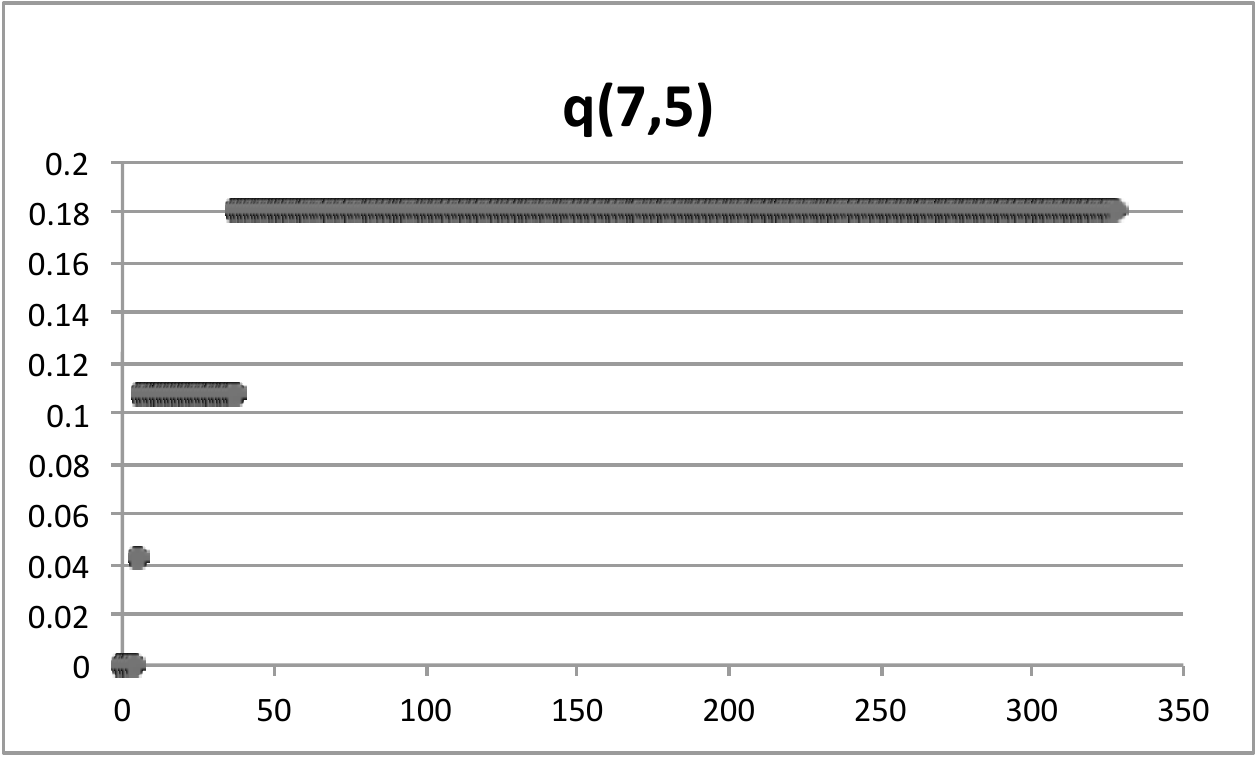}
	\end{subfigure}
	\begin{subfigure} 
		\centering
		\includegraphics[width=2.7cm]{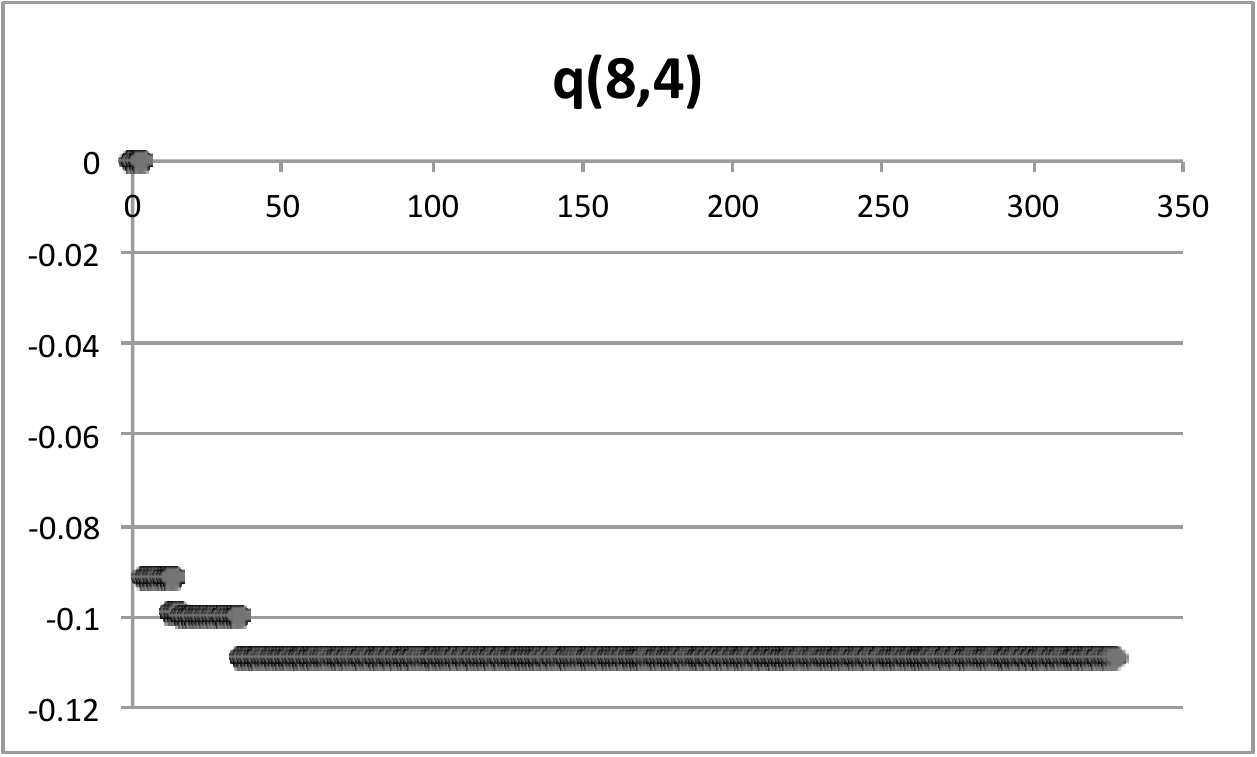}
	\end{subfigure}
	\vfill
	\begin{subfigure} 
		\centering
		\includegraphics[width=2.8cm]{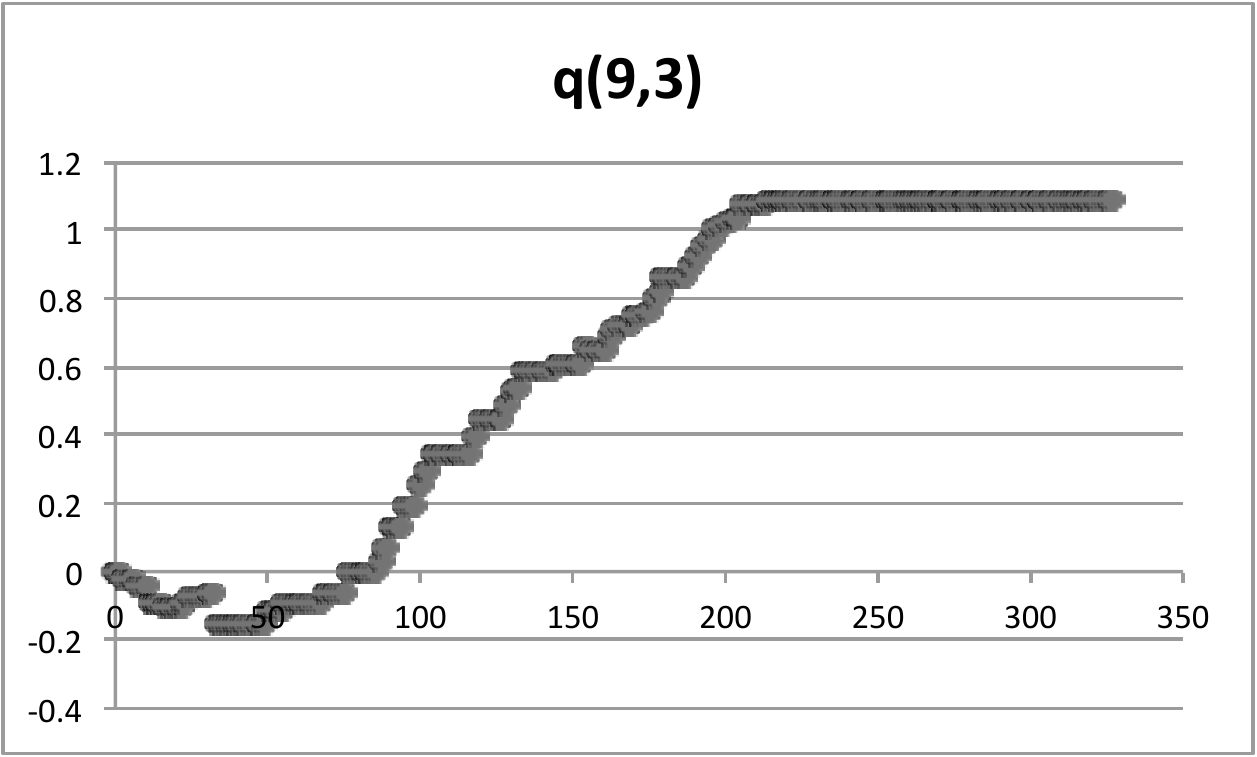}
	\end{subfigure}
	\begin{subfigure} 
		\centering
		\includegraphics[width=2.8cm]{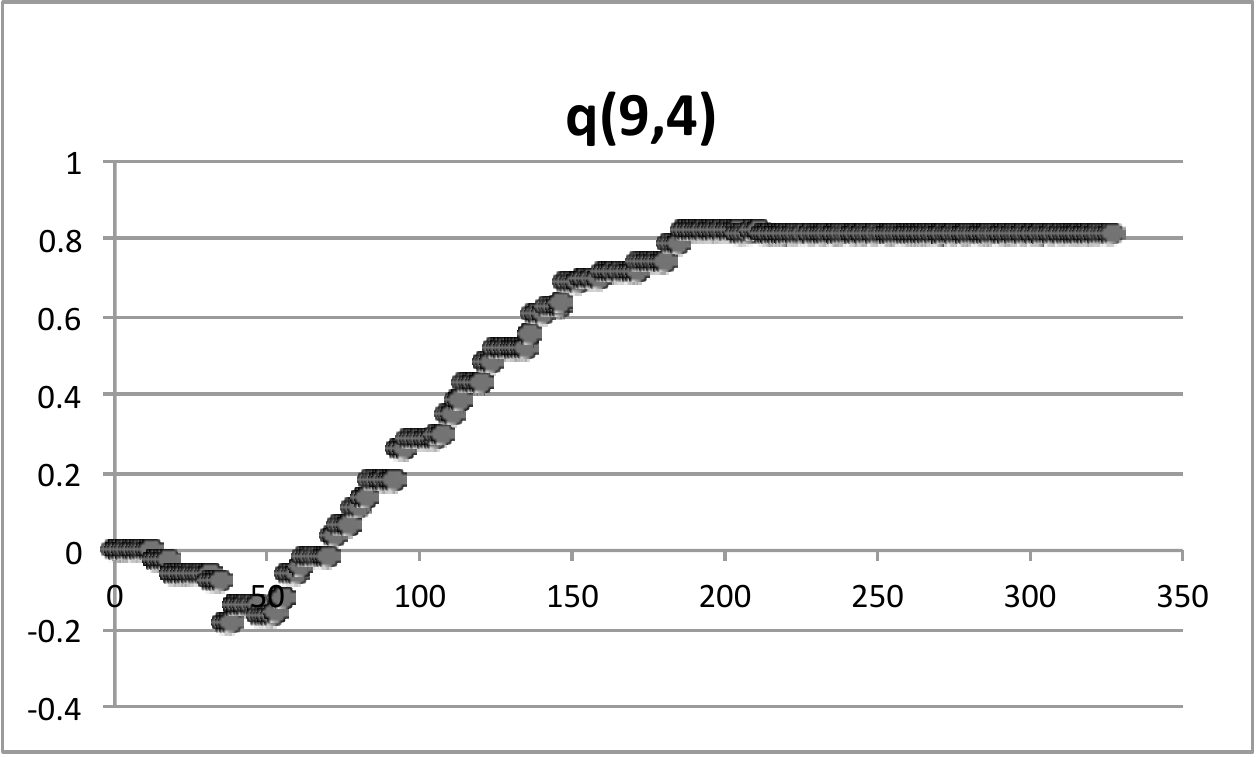}
	\end{subfigure}
	\begin{subfigure} 
		\centering
		\includegraphics[width=2.7cm]{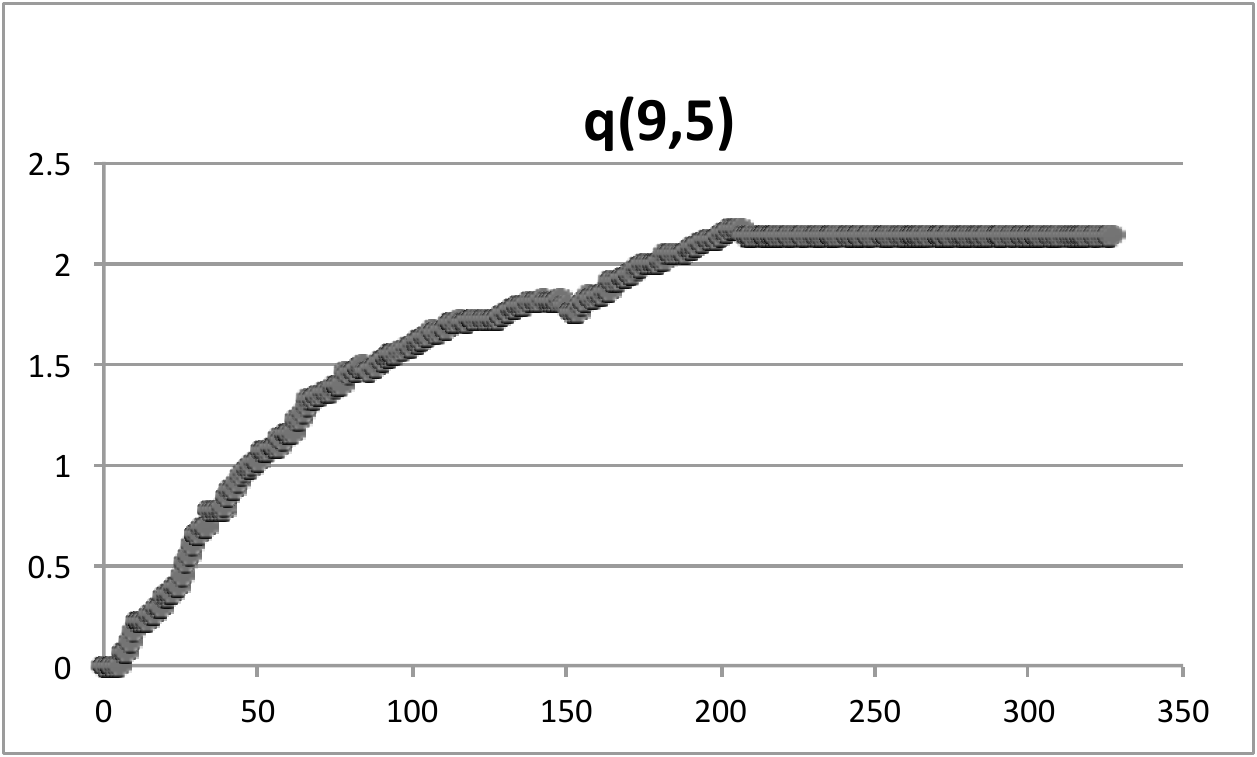}
	\end{subfigure}

	\caption{Temporal evolution of q-values.}
	\label{fig:qvalues}
\end{figure}

The temporary evolution of the q-values associated to each state-action pairs for the learning strategy $S1$ is shown (for partial set of pairs) in Figure \ref{fig:qvalues}. Note that the change in the q-values occurs when the corresponding rule is activated, i.e., when the system is in state $S(t)$ and takes specific action $a_i$. As the figure shows, some q-values changed to a negative value during exploration phase. It means that these actions are basically punished and as a result are not appropriate to be taken in the future. The optimal consequent for each rule in the rule base is determined by the most highest q-value at the end of the learning phase. For instance, action $a_5$ is the best consequent for rule number $9$ in learning strategy $S1$. 

\begin{figure}
	\centering
	\begin{subfigure} 
			\centering
			\includegraphics[width=4cm]{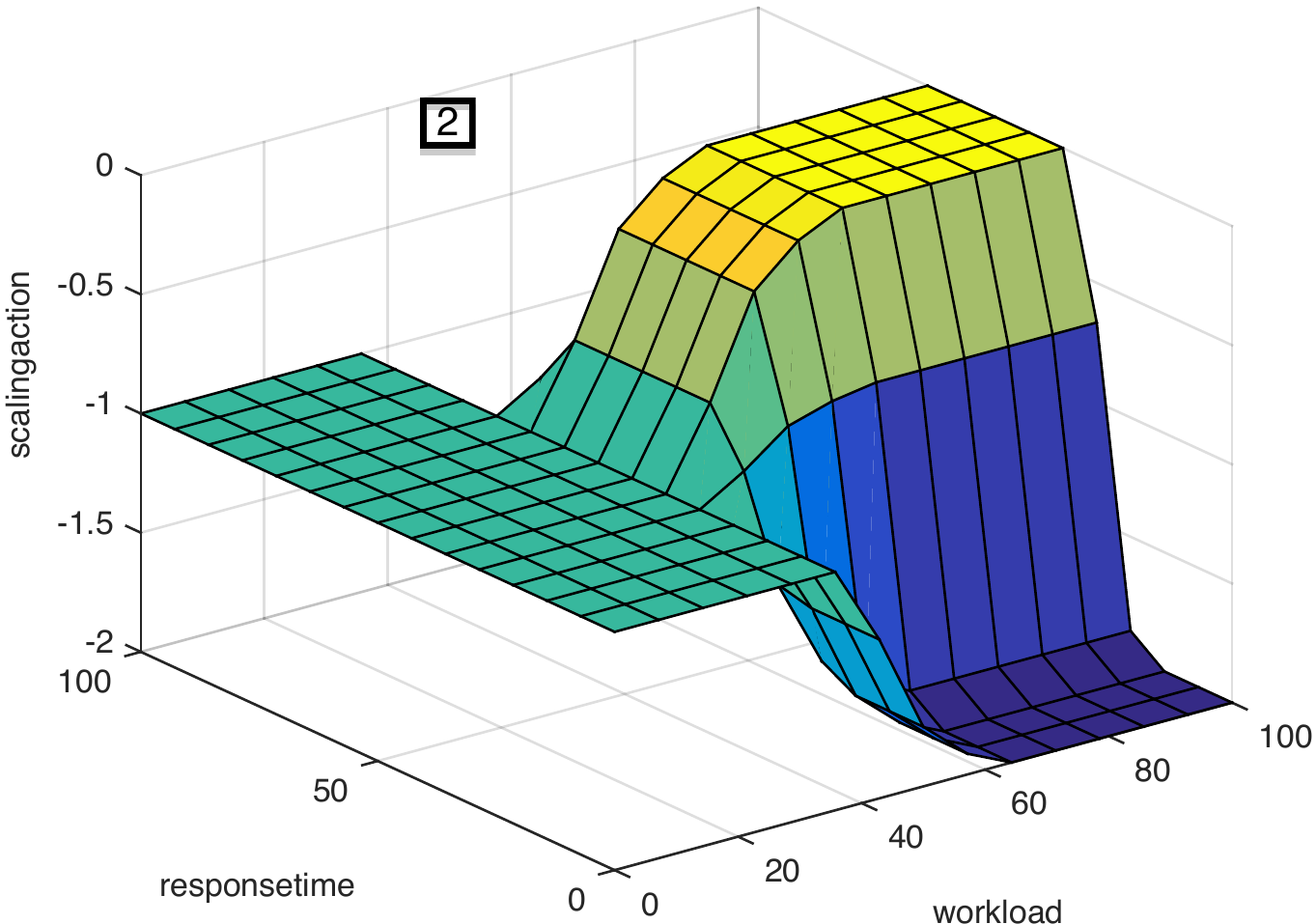}
	\end{subfigure}		
\quad
	\begin{subfigure}
		\centering
		\includegraphics[width=4cm]{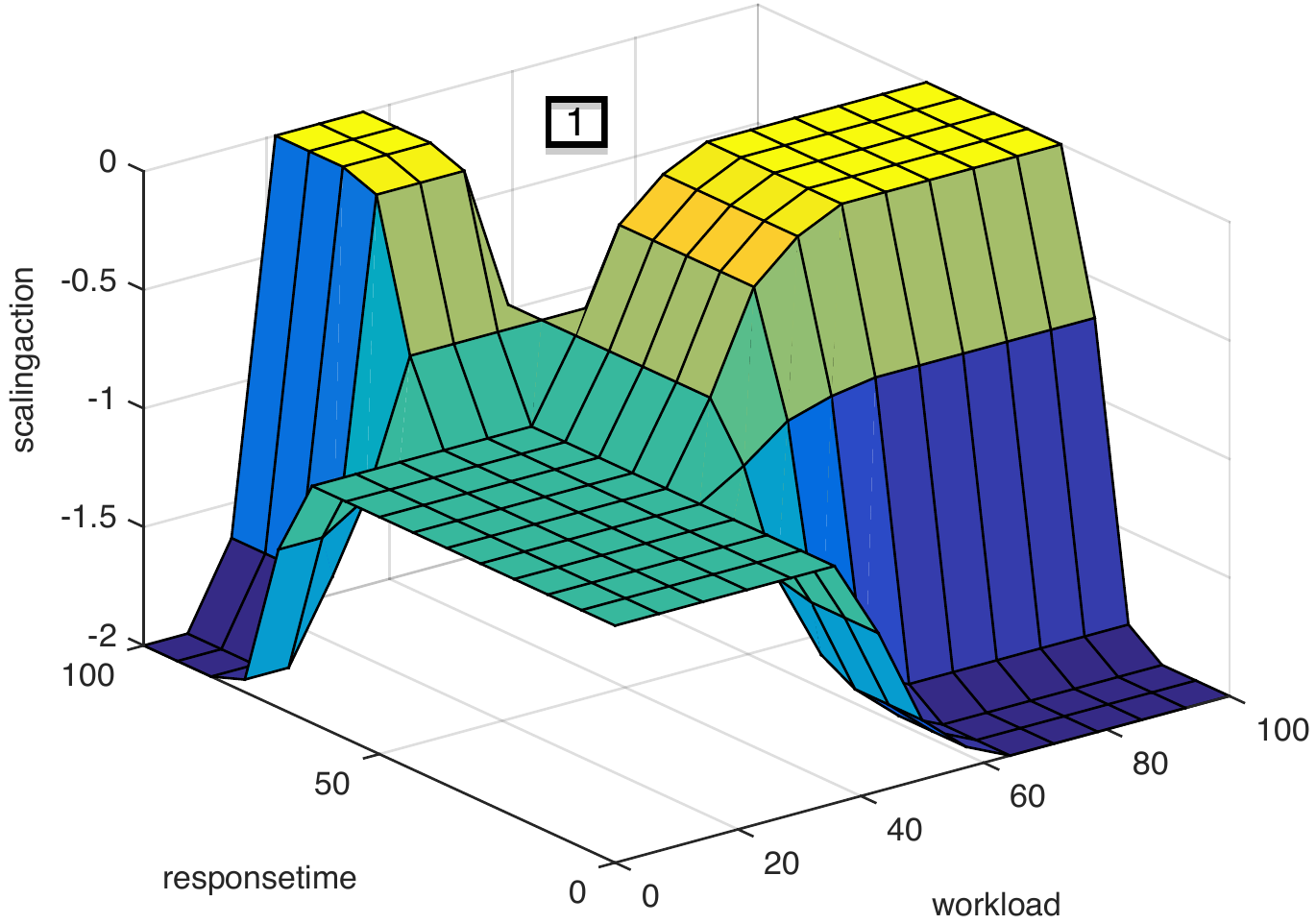}
	\end{subfigure}
\vfil
	\begin{subfigure}
				\centering
				\includegraphics[width=4cm]{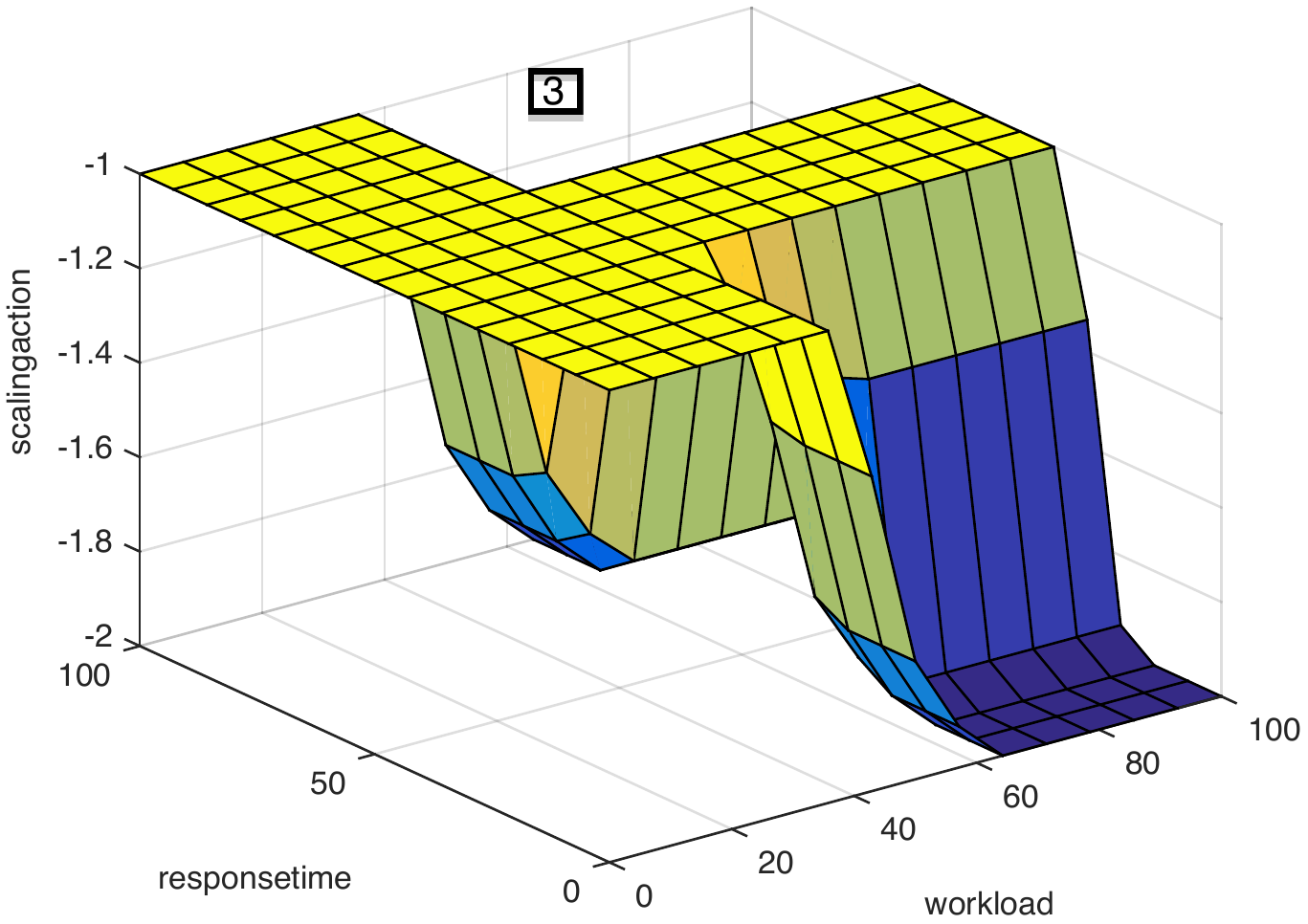}
	\end{subfigure}
		\quad
	\begin{subfigure}
			\centering
			\includegraphics[width=4cm]{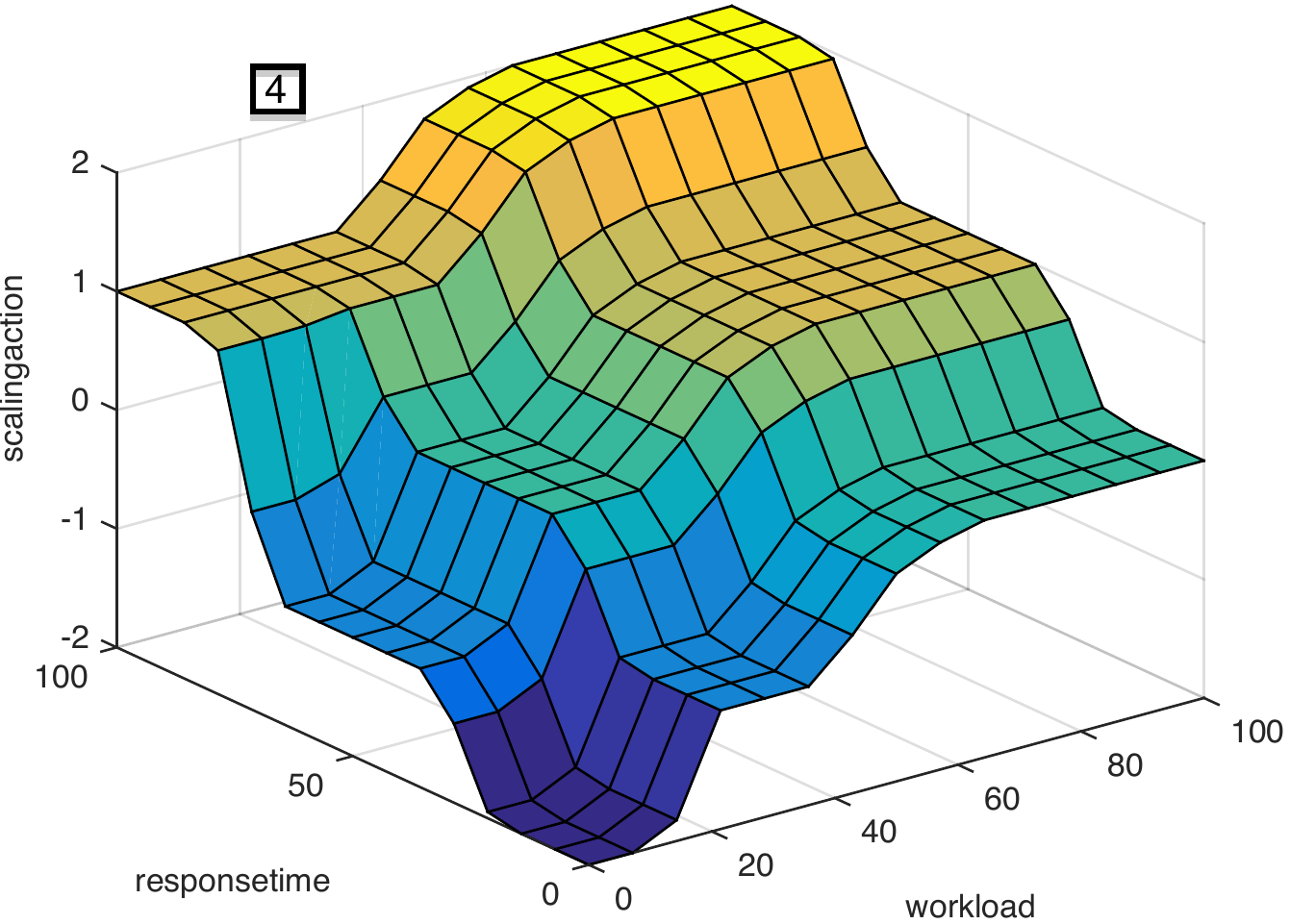}
	\end{subfigure}
	\caption{Temporal evolution of control surface.}
	\label{fig:surfaces}
\end{figure}

In accordance to the change in the q-values, the control surface of the fuzzy controller is also evolving. Figure \ref{fig:surfaces} shows the temporal evolution in control surface of the fuzzy controller. The initial design-time surface is not shown as it is a constant plane at point zero. The surface is evolved until the learning has been converged. Note that the first surface is the one in the upper left, then upper right, lower left and the final surface is located at the lower right corner when the learning has been converged.

\begin{figure}
	\begin{center}
		\includegraphics[width=6cm]{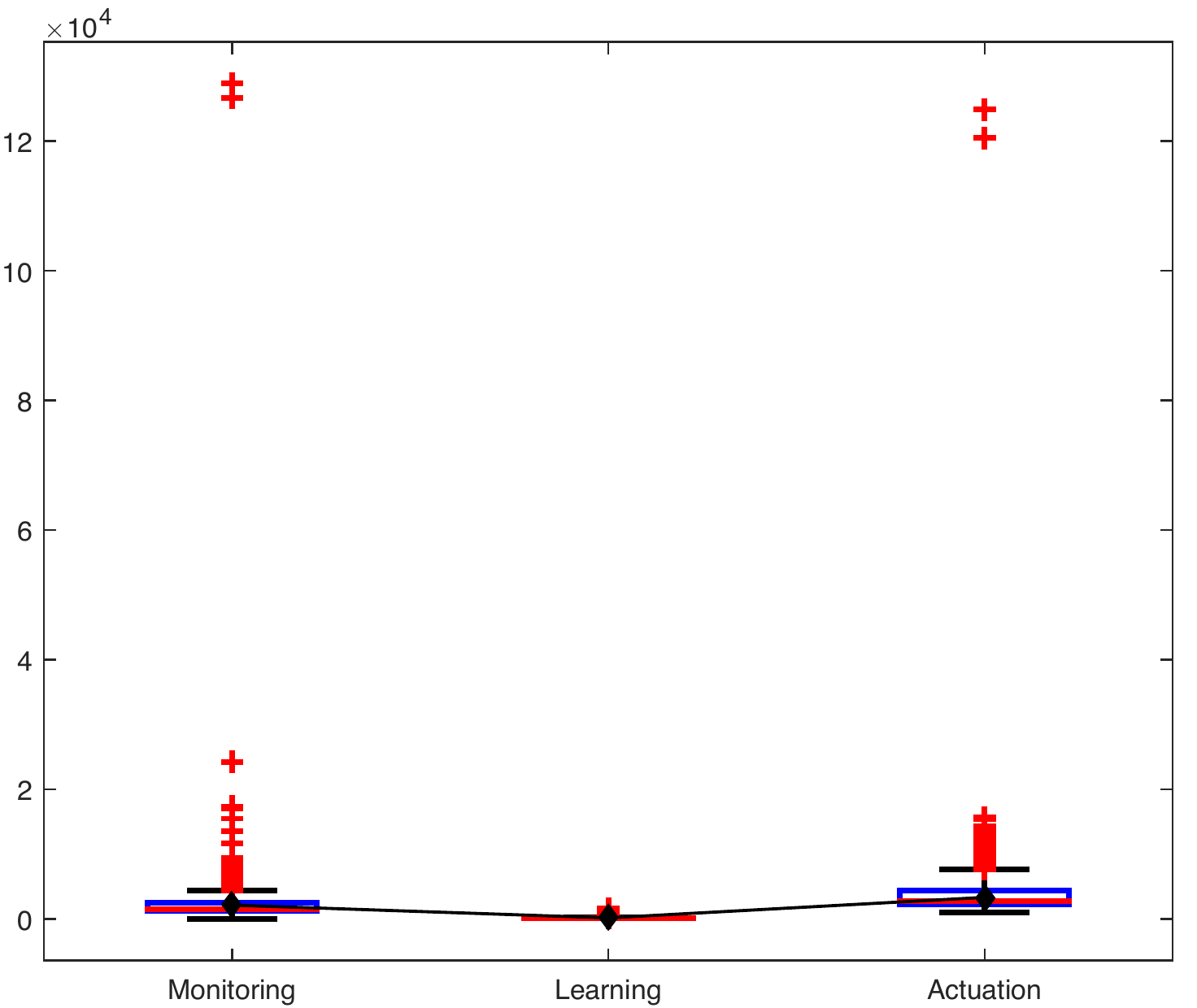}
		\caption{Runtime delay for MAPE loop activities.}
		\label{fig:runtimeboxes}
	\end{center}
\end{figure}

The runtime overhead of the feedback control loop activities (cf. Figure \ref{fig:implementation}) is depicted in Figure \ref{fig:runtimeboxes}. We collected these data points in each control loop (i.e., more than $8600$ data points). As the result shows, the learning overhead is in the order of $100ms$ and the monitoring and actuation delay is in the order of $1000ms$. Note that the actuation delay is only measured for issuing the change command and it does not include the enaction time as it is in the order of several minutes.



\subsection{FQL4KE Flexibility and Effectiveness (RQ2)}
\label{sec:rq2}
\begin{figure}
	\begin{center}
		\includegraphics[width=8cm]{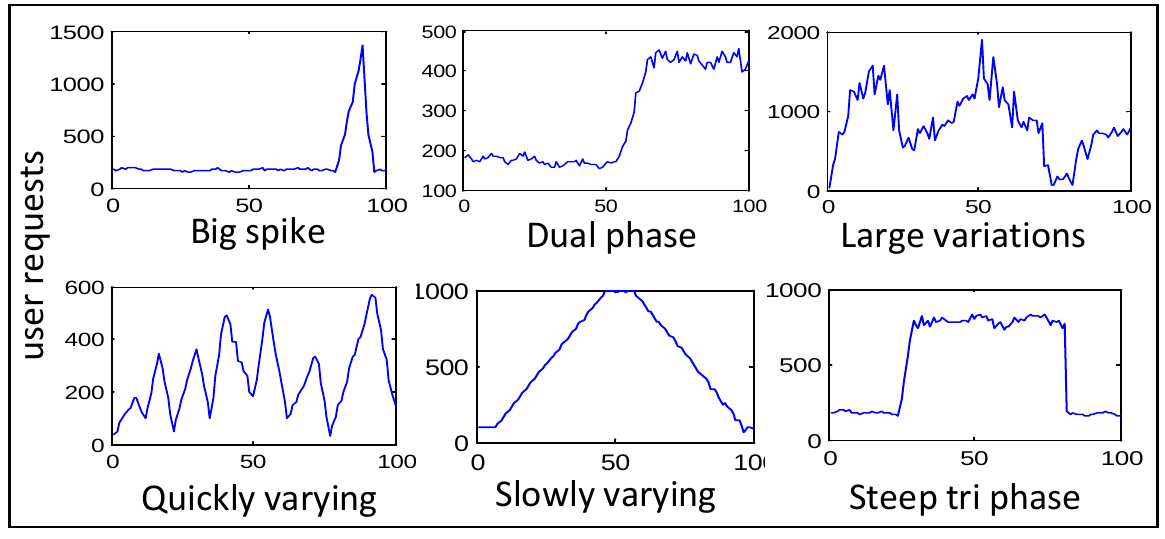}
		\caption{Synthetic workload patterns.}
		\label{fig:workloads}
	\end{center}
\end{figure}

Let us now investigate how effective the self-learning mechanism is in \texttt{FQL4KE}. More specifically, we want to study how the learning component of \texttt{FQL4KE} improves the functionality of dynamic resource allocation over static rule-based or native mechanisms. Table \ref{tab:comparison} summarizes the criteria that we considered for comparing different auto-scaling strategies with respect to different workload patterns. Note that strategies $S5$ corresponds to the fuzzy controller with initial knowledge extracted from users at design-time with no learning component and the last strategy corresponds to Azure native auto-scaling. We synthetically generated 6 different workload patterns (see Figure \ref{fig:workloads}) in order to provide enough environmental conditions for this comparison. The $x$ axis shows the experimental time and the $y$ axis shows the number (in $[0,100]$) for which the Fibonacci series needs to be calculated, demonstrating the workload intensity similar to the number of concurrent users for a web-based application. A key parameter in learning-based approaches is the convergence delay to reach the optimal policy. The response time of the system under different workloads is also considered as another comparison criterion. The average number of VMs acquired throughout the experiment interval as well as the number of changes in the underlying resources (i.e., sum of issued scaling actions) is also considered as a comparison criterion. 
The main findings described in Table \ref{tab:comparison} can be summarized as follows: 
\begin{itemize}
	\item Sequential decreasing of exploration factor (cf. $S1$) is effective in accelerating learning convergence. However, it is also effective for highly dynamic workloads such as "quickly varying" as in Figure \ref{fig:workloads} because it keeps a minimum of $\epsilon=0.2$ when initial knowledge has been learned, it then helps to keep the rules updated when new situations arise.
	\item Initial high exploration (cf. $S2$) is effective for quick convergence, but in non-predictable workloads such as ''quickly varying", the decisions become sub-optimal. This is evident by comparing the average number of VMs and the number of learning iteration until convergence for "large variation" and "quickly varying" patterns. 
	\item Although high constant exploration (cf. $S3$) is effective in unpredictable environment (see response time and compare it with other strategies), it is not optimal in terms of convergence, number of changes and acquired resources. Note that the higher number of changes in the resources means that for quite considerable period in time there exists some instability in the deployment environment of the application.
	\item Maximum exploration rate (cf. $S4$) is not a good learning strategy by no means as it only produces random actions and it never converges to an optimal policy.
	\item The strategy $S5$ is \texttt{RobusT2Scale} cloud controller, representing a policy-based adaptation without any policy adaptation process. By comparing response time, number of changes and average number of resources (almost in all aspects and for all patterns it is relatively lower), we can observe that \texttt{FQL4KE} is effective in terms of learning optimal policies and updating them at runtime. 
	\item Both the cloud controller without learning mechanism and with learning are more effective than the native cloud platform reactive auto-scalers. Note that for the controller without learning, we consider a reasonably logical set of rules to govern the elasticity decision making. But if we consider a non sensible set of rules, the native auto-scaling of Azure would beat \texttt{RobusT2Scale}.
\end{itemize}

\begin{table}[htbp]
	\centering
	\caption{Comparison of the effectiveness of exploration/exploitation strategies under different workloads.}{
		\centering
		\scriptsize
		\begin{tabular}{crccc}
			\toprule
			\textbf{Strategy} & \multicolumn{1}{c}{\textbf{Criteria}} & \textbf{Big spike} & \textbf{Dual phase} & \textbf{Large variations} \\
			\midrule
			\multirow{2}[4]{*}{\textbf{S1}} &   $rt_{95\%}, \overline{vm}$ & 1212ms, 2.2 & 548ms, 3.6 & 991ms, 4.3 \\
			& \multicolumn{1}{c}{node change} & 390     & 360     & 420 \\
			& \multicolumn{1}{c}{convergence} & 32     & 34     & 40 \\
			\multirow{2}[4]{*}{\textbf{S2}} &   $rt_{95\%}, \overline{vm}$    & 1298ms, 2.3 & 609ms, 3.8 & 1191ms, 4.4 \\
			& \multicolumn{1}{c}{node change} & 412     & 376     & 429 \\
			& \multicolumn{1}{c}{convergence} & 38     & 36     & 87 \\
			\multirow{2}[4]{*}{\textbf{S3}} &   $rt_{95\%}, \overline{vm}$    & 1262ms, 2.4 & 701ms, 3.8 & 1203ms, 4.3 \\
			& \multicolumn{1}{c}{node change} & 420     & 387     & 432 \\
			& \multicolumn{1}{c}{convergence} & 30     & 29     & 68 \\
			\multirow{2}[4]{*}{\textbf{S4}} &   $rt_{95\%}, \overline{vm}$    & 1193ms, 3.2 & 723ms, 4.1 & 1594ms, 4.8 \\
			& \multicolumn{1}{c}{node change} & 487     & 421     & 453 \\
			& \multicolumn{1}{c}{convergence} & 328     & 328     & 328 \\
			\multirow{2}[4]{*}{\textbf{S5}} &   $rt_{95\%}, \overline{vm}$    & 1339ms, 3.2 & 729ms, 3.8 & 1233ms, 5.1 \\
			& \multicolumn{1}{c}{node change} & 410     & 377     & 420 \\
			& \multicolumn{1}{c}{convergence} & N/A     & N/A     & N/A \\
			\multirow{2}[4]{*}{\textbf{Azure}} &   $rt_{95\%}, \overline{vm}$    & 1409ms, 3.3 & 712ms, 4.0 & 1341ms, 5.5 \\
			& \multicolumn{1}{c}{node change} & 330     & 299     & 367 \\
			& \multicolumn{1}{c}{convergence} & N/A     & N/A     & N/A \\
			\bottomrule
			\textbf{} & \multicolumn{1}{c}{\textbf{}} & \textbf{Quickly varying} & \textbf{Slowly varying} & \textbf{Steep tri phase} \\
			\midrule
			\multirow{2}[4]{*}{\textbf{S1}} &   $rt_{95\%}, \overline{vm}$    & 1319ms, 4.4 & 512ms, 3.6 & 561ms, 3.4 \\
			& \multicolumn{1}{c}{node change} & 432    & 355     & 375 \\
			& \multicolumn{1}{c}{convergence} & 65     & 24     & 27 \\
			\multirow{2}[4]{*}{\textbf{S2}} &    $rt_{95\%}, \overline{vm}$   & 1350ms, 4.8 & 533ms, 3.6 & 603ms, 3.4 \\
			& \multicolumn{1}{c}{node change} & 486    & 370     & 393 \\
			& \multicolumn{1}{c}{convergence} & 98     & 45     & 28 \\
			\multirow{2}[4]{*}{\textbf{S3}} &    $rt_{95\%}, \overline{vm}$   & 1287ms, 4.9 & 507ms, 3.7 & 569ms, 3.4 \\
			& \multicolumn{1}{c}{node change} & 512    & 372     & 412 \\
			& \multicolumn{1}{c}{convergence} & 86     & 40     & 23 \\
			\multirow{2}[4]{*}{\textbf{S4}} &    $rt_{95\%}, \overline{vm}$   & 2098ms, 5.9 & 572ms, 5.0 & 722ms, 4.8 \\
			& \multicolumn{1}{c}{node change} & 542    & 411     & 444 \\
			& \multicolumn{1}{c}{convergence} & 328     & 328     & 328 \\
			\multirow{2}[4]{*}{\textbf{S5}} &    $rt_{95\%}, \overline{vm}$   & 1341ms, 5.3 & 567ms, 3.7 & 512ms, 3.9 \\
			& \multicolumn{1}{c}{node change} & 479    & 366     & 390 \\
			& \multicolumn{1}{c}{convergence} & N/A     & N/A     & N/A \\
			\multirow{2}[4]{*}{\textbf{Azure}} &   $rt_{95\%}, \overline{vm}$    & 1431ms, 5.4 & 1101ms, 3.7 & 1412ms, 4.0 \\
			& \multicolumn{1}{c}{node change} & 398     & 287     & 231 \\
			& \multicolumn{1}{c}{convergence} & N/A    & N/A     & N/A \\			
			\bottomrule
		\end{tabular}}%
		\label{tab:comparison}%
	\end{table}%





%
%
%




\section{discussion}
\label{sec:discussion}



\subsection{Computational Complexity and Memory Consumption}

Step 2 to Step 8 in Algorithm \ref{alg:fql} are the main computationally intensive steps of our approach and based on our experiments are in the order of few minutes for 9 states and for 10,000 learning epochs. However, this is not an issue in our setting because the control loops are long enough due to the fact that scaling out actions are also in the order of magnitude of several minutes, 8-9 minutes for scaling out an extra small VM on Azure platform and 2-3 minutes for removing an existing VM.

In addition, the memory consumption of our approach is given by the dimensions of the look up table that saves and updates the q-values. In other words, the space complexity of our approach is always $O(N\times J)$, where $N$ is the number of states and $A$ is the number of actions. In the setting that we described in this paper, the table is composed by $9$ states $\times 5$ actions $= 45$ q-values, thus memory consumption is negligible.  

%

\subsection{FQL4KE for Policy-based Adaptations}
Although in this paper we showed applicability of \texttt{FQL4KE} with \texttt{RobusT2Scale}, this approach is general and can be integrated with any knowledge-based controllers. By knowledge-based controller, we mean any controller that have explicit notion of knowledge that can be specified in terms of rules and used for reasoning and producing the control signal. Basically \texttt{FQL4KE} can be integrated with such controllers to learn rules and populate the knowledge base at runtime. Such policy-based controllers are not only applied for resource scaling but have also been previously applied to the rule-based adaptations of software architecture at runtime \cite{Georgas2008,Georgas2009}. 

\subsection{Limitations of FQL4KE}
Besides the provided features of \texttt{FQL4KE}, it comes with some limitations. Firstly, performance of scaling actions produced by \texttt{FQL4KE} during initial learning epochs at runtime may be poor. This imposes some difficulties. First, at early stages when the learning process has not been converged there might be some over-provisioning or under-provisioning due to such decisions. However, some other strategies (e.g., temporary over-provisioning) can be adopted in parallel in order to let the approach learns policies and after it learned optimal policies, it becomes the sole decision maker for resource allocation. Secondly, the learning process may be sensitive to the selection of the reinforcement signal (cf. Equation \ref{eq:utility-function}). It is also dependent on the fact that the system states must have been visited sufficiently \cite{sutton1998introduction}. 

\subsection{Threats to Validity}

There are a number of sources of threats to the validity of the results presented in Section \ref{sec:experiment}. 
First, the results presented in Table \ref{tab:comparison} may be slightly different depending on the utility function defined in Eq. \ref{eq:utility-function}. We defined a reasonable function to measure the reward, while this can be defined differently leading to a different effectiveness of learning strategies. We expect the results would be consistent with the effectiveness (cf. Table \ref{tab:comparison}) of our solution as long as the function is appropriate, i.e., only consider both reward or punishment even with different metrics that we used, but not only one aspect. 

The other threat to the validity of the result is the application framework that we built for our experiment, i.e., \texttt{ElasticBench}. Although we embed different characteristics of a cloud-based application by using Fibonacci based calculation and using cloud based technologies such as caching, but the results presented in Table~\ref{tab:comparison} may be slightly different for another type of application. However, since we can simulate different functionalities with this application framework, we expect that results on a different application is consistent with the ones presented in Section \ref{sec:experiment}. This requires further investigations with real-world software applications. 

Although the approach does not impose any constraints on the possible number of scaling actions, for simplicity we only consider five possible scaling actions (i.e., ${-2,-1,0,+2,+2}$) for describing the approach and evaluations. This limited set of actions has some implications on the performance (cf. Section \ref{sec:rq1}) and effectiveness of learning (cf. Section \ref{sec:rq2}).

Finally, limited number of workload patterns (6 patterns is used in this work for evaluation, cf. Figure \ref{fig:workloads}) is another threats to the validity. As it is also used in other research \cite{gandhi2014adaptive}, this set of patterns, although not comprehensive, but provides a reasonably enough environmental conditions for evaluation.

%

\section{Related Work}
\label{sec:related}

In autonomic computing \cite{Kephart2003}, exploiting policy-based adaptation techniques to build self-adaptive software has been attractive. In the following, instead of reviewing auto-scaling approaches which has been comprehensively reviewed from different aspects in \cite{Lorido-Botran2014}\cite{caron2012auto}\cite{Galante2012}\cite{gambi2013assurance}\cite{Ardagna2014}, we only consider related work to fundamental attributes of autonomic computing (i.e., self-adaptiveness, self-learning, self-optimizing). We then single out and categorize the selected works whose focus is on \textit{policy-based adaptation}, whether it is related to \textit{software adaptation} or \textit{dynamic resource allocation}. 

\emph{Policy-based adaptation.} In self-adaptive software literature, policy-based adaptation has gained momentum due to its efficiency and flexibility for adaptation planning \cite{Ho2015}. Policy, in general, is a mapping between a situation or condition, to appropriate action, strategy or adaptation. A policy-based approach can potentially decouple adaptation logic with how to react when necessary. Rainbow \cite{Garlan2004} exploits architecture-based adaptation, in which system chooses new architectural reconfiguration, at runtime, based on rules defined at design-time. In a similar line, Sykes et al. \cite{Sykes2008} propose an online planning approach to architecture-based self-managed systems. Their work describes plan, as a set of condition-action rules, which has been generated by observing a change in the operational environment or a system failure. Georgas and Taylor \cite{Georgas2004} present a architecture-centric knowledge-based approach in which adaptation polices are specified as reaction rules. Not all of the policy-based approaches exploit if-then rules, other resemblances of policy have been also utilized. For instance, model-based approaches in terms of variability models has been adopted in \cite{Elkhodary2010}. While policy-based approaches have been shown useful in some settings (e.g., enforcing certain characteristics in the system), they cannot deal with unseen situations or uncertainties. System hence produces suboptimal decision, as there is no automatic mechanism to react when exceptions occur and usually need human intervention \cite{Ho2015}. To address the issues, online policy evolution has been introduced \cite{Kim09, Abbas2011}. The solution proposed in this paper, \texttt{FQL4KE}, is in the same line of research, but applied fuzzy Q-learning, for the first time, to the problem of dynamic resource allocation. 

\emph{Dynamic Adaptation Planning.} Qian et al. \cite{John14} exploits case-based reasoning to improve the effectiveness of adaptation planning by learning from past experiences. Goal models are used to represent system requirements of self-adaptive systems, while the adaptation is essentially a search to find a match between new situations and the closest prior cases of adaptation. In \cite{Bencomo13}, dynamic decision networks are proposed to deal with the uncertainty in decision-making of self-adaptive systems. The initial models are provided by experts; however, the models are updated at runtime as more evidences are observed through monitoring. Esfahani et al. \cite{Esfahani13} discuss the application of black-box learning models to understand the impact of different features in a self-adaptive system. Given a system goal, a function is learned to formulate the impact of different features, and accordingly the features are enabled or disabled to adapt to the changes and achieve the goals. Amoui et al. \cite{Amoui08} present an approach based on reinforcement learning to select adaptation actions at runtime. Through an adaptive web-based case study, it is shown that the approach provides similar results comparing to a voting-based approach that uses expert knowledge. Kim et al. \cite{Kim09} discuss the application of Q-learning to plan architecture-based adaptations, a similar policy-based architecture adaptation is also proposed in \cite{Georgas2008}. These approaches are applied in robotics domain. Similarly to these works, \texttt{FQL4KE} is proposed to address the issue of decision making in autonomic systems; however, it particularly focuses on resource allocation in cloud-based applications.

\emph{Dynamic Resource Allocation.} Xu et al. \cite{URL,URLCoordinated} present an approach to learning appropriate auto-configuration in virtualized resources. It uses multiple agents, each of which apply reinforcement learning to optimize auto-configuration of its dedicated environment. Barrett et al. \cite{Barrett13} investigate the impact of varying performance of cloud resources on application performance. They show that a resource allocation approach, considering this aspect, achieves benefits in terms of performance and cost. To reduce the learning time, a parallelized reinforcement learning algorithm is proposed through which multiple agents are employed to deal with the same tasks to speed up the procedure to explore the state space. The reward values are calculated by combining the accumulated experience of different agents. In a similar approach \cite{Dutreilh2011} appropriate initialization of the q-values is proposed to accelerate the learning convergence.
Tesauro et al. \cite{Tesauro2006} demonstrate how to combine the strengths of both RL (model-free) and queuing models (model-based) in a hybrid approach, in which their RL needs to be trained at design-time while at runtime a queuing model policy controls the system. In \cite{DynaQoS}, a multi-layer approach is presented to handle multi-objective requirements such as performance and power in dynamic resource allocation. The lower layer focuses on each objective, and exploits a fuzzy controller proposed earlier in \cite{Xu07}. The higher layer is to maintain a trade-off by coordinating the controllers. Lama et al. \cite{Lama2013} integrate NN with fuzzy logic to build adaptive controllers for autonomic server provisioning. Similar to our approach, NNs define a set of fuzzy rules, and the self-adaptive controller adapts the structure of the NN at runtime, therefore automatically updating rules. Unlike the above approaches, \texttt{FQL4KE} offers a seamless knowledge evolution through fuzzy control and RL, putting aside the burden that was on the shoulder of users. 

\section{Conclusion and Future Work}
\label{sec:future}

This paper has investigated the notion of knowledge evolution in dynamic resource provisioning for cloud-based applications. The scenario under study assumes no a priori knowledge is available regarding elasticity policies that cloud controllers can exploit. More precisely, instead of specifying elasticity policies as a typical case in auto-scaling solutions, system operators are now only required to provide the importance weights in reward functions. In order to realize this, a fuzzy rule-based controller (the lower feedback control loop in Figure \ref{fig:knowledge-evolution}) entangled with a reinforcement learning algorithm (the upper knowledge evolution loop in Figure \ref{fig:knowledge-evolution}) for learning optimal elasticity policies, has been proposed. The main advantages of the proposed approach are as follows:
\begin{enumerate}
	\item \texttt{FQL4KE} is \emph{robust} to highly dynamic workload intensity due to its self-adaptive and self-learning capabilities.
	\item \texttt{FQL4KE} is \emph{model-independent}. The variations in the performance of the deployed applications and the unpredictability of dynamic workloads do not affect the effectiveness of the proposed approach.
	\item \texttt{FQL4KE} is capable of automatically constructing the control rules and keeping control parameters updated through \emph{fast online learning}. It executes resource allocation and learns to improve its performance simultaneously.
	\item Unlike supervised techniques that learn from the training data, \texttt{FQL4KE} \emph{does not require off-line training} that saves significant amount of time and efforts.
\end{enumerate}


We plan to extend our approach in a number of ways: (i) extending \texttt{FQL4KE} to perform in the environments where only partially observable (for this we will exploit partially observable Markov decision processes), (ii) exploiting clustering approaches to learn the membership functions of the antecedents (in this work we assume they will not change once they specified, for enabling the dynamic change we will consider incremental clustering approaches) in fuzzy rules.


\section*{Acknowledgment}

The authors would like to thank Soodeh Farokhi and Saeid Masoumzadeh for their constructive comments on the final draft of the paper. The research leading to these results has received funding from IC4 (an Irish national technology centre funded by Enterprise Ireland) and the EU's Programme FP7/2007-2013 under grant agreement 610802 (CloudWave).



%

\bibliographystyle{plain}
\bibliography{cac15}

\end{document}